\documentclass[11pt,a4paper]{article}
\usepackage{amsfonts,amssymb, amsmath, amsthm, stmaryrd, multibox, hyperref}

\def\mathun{\hbox{ 1\hskip -3pt l}}

\begin{document}

\begin{center}{\Large \textbf{
The unitary representations of the Poincar\'e group in any spacetime dimension}}\end{center}

\hspace{.4cm}

\begin{center}
{\large{Xavier Bekaert \textsuperscript{a} ~and~ 
Nicolas Boulanger \textsuperscript{b}}}
\end{center}

\hspace{.4cm}

\begin{center}
$^{a}$ Institut Denis Poisson, Unit\'e mixte de Recherche 7013, 
Universit\'e de Tours, Universit\'e d'Orl\'eans, CNRS,
Parc de Grandmont, 37200 Tours (France)\\
{\tt xavier.bekaert@lmpt.univ-tours.fr}
\\

\hspace{.4cm}

$^{b}$ Service de Physique de l'Univers, Champs et Gravitation\\
Universit\'e de Mons -- UMONS, 
Place du Parc 20, 7000 Mons (Belgium)
\\
{\tt nicolas.boulanger@umons.ac.be}
\end{center}



\hspace{3cm}

An extensive group-theoretical treatment of linear relativistic field
equations on Minkowski spacetime of arbitrary dimension $D>2$ is presented
in these lecture notes. To start with, the one-to-one correspondence
between linear relativistic field equations and unitary representations of
the isometry group is reviewed. In turn, the method of induced
representations reduces the problem of classifying the representations of
the Poincar\'e group $ISO(D-1,1)$ to the classification of the representations
of the stability subgroups only. Therefore, an exhaustive treatment of the
two most important classes of unitary irreducible representations,
corresponding to massive and massless particles (the latter class
decomposing in turn into the ``helicity'' and the ``infinite-spin''
representations) may be performed via the well-known representation theory
of the orthogonal groups $O(n)$ (with $D-4<n<D\,$). Finally, covariant field
equations are given for each unitary irreducible representation of the
Poincar\'e group with non-negative mass-squared. Tachyonic representations
are also examined. All these steps are covered in many details and with
examples. The present notes also include a self-contained review of the
representation theory of the general linear and (in)homogeneous orthogonal
groups in terms of Young diagrams.
\vspace*{3mm}


\pagebreak\
\tableofcontents

\section{Group-theoretical preliminaries}
\label{sec:intro}

Elementary knowledge of the theory of Lie groups and their
representations is assumed (see \textit{e.g.} the textbooks
\cite{BeBo:BHall,BeBo:Barut} or the lecture notes
\cite{BeBo:Liealgs}). The basic definitions of the Lorentz and
Poincar\'e groups together with some general facts on the theory
of unitary representations are reviewed in order to fix the 
notation and settle down the prerequisites.

\subsection{Universal covering of the Lorentz group}
\label{BeBo:Lorentzgr}
%
The group of linear homogeneous transformations
$x'^{\mu}=\Lambda^{\mu}_{~\nu}x^{\nu}$ ($\mu,\nu=0,1,\ldots,D-1$)
preserving the Minkowski metric $\eta_{\mu\nu}$ of
``mostly plus" signature
$(-,+,\ldots,+)\,$, $$\Lambda^T\eta\Lambda=\eta\,,$$ where
$\Lambda^T$ denotes the matrix transpose of $\Lambda\,$, is called
the \textsl{Lorentz group} $O(D-1,1)$.

A massless particle propagates on the light-cone $x^2=0\,$. 
Without loss of generality, one may consider that its 
momentum points along the $(D-1)$th spatial direction. Then it turns
out to be convenient to make use of the \textsl{light-cone
coordinates} $$x^{\pm}=\frac{1}{\sqrt{2}}\,(\,x^{D-1}\,\pm\,
x^0\,)\quad \mbox{and}\quad x^m\quad (m=1,\ldots,D-2)\,,$$ where
the Minkowski metric reads $\eta_{++}=0=\eta_{--}\,$,
$\eta_{+-}=1\,=\eta_{-+}\,$ and $\eta_{mn}=\delta_{mn}$
($m,n=1,\ldots,D-2$).

On physical grounds, one will mainly be interested in the 
matrices $\Lambda$'s with determinant $+1\,$ and such that 
$\Lambda^0_{~0}\geqslant 0\,$. It can be
shown that such matrices $\Lambda$'s also form a group that one
calls the \textsl{proper orthochronous Lorentz group} denoted by
$SO(D-1,1)^{\uparrow}\,$. It is connected to the identity, but not
{\textsl{simply connected}}, that is to say, there exist loops in
the group manifold $SO(D-1,1)^{\uparrow}$ which are not
continuously contractible to a point. In order to study the
representations (reps) of $SO(D-1,1)^{\uparrow}\,$, one may first
determine its universal covering group, \textit{i.e.} the Lie
group which is simply connected and whose Lie algebra is
isomorphic to $\mathfrak{so}(D-1,1)\,$, the Lie algebra of
$SO(D-1,1)^{\uparrow}\,$. For $D\geqslant 4\,$, the universal
covering group, denoted $Spin(D-1,1)\,$, is the double cover of
$SO(D-1,1)^{\uparrow}\,$. The spin groups $Spin(D-1,1)$ have no
intrinsically projective representations. Therefore, a single (or
double) valued ``representation" of $SO(D-1,1)^{\uparrow}$ is
meant to be a genuine representation of $Spin(D-1,1)\,$.

\vspace{1mm}\noindent\textbf{Warning:} The double cover of
$SO(2,1)^\uparrow$ is the group $SU(1,1)$, in close analogy to the
fact that the double cover of $SO(3)$ is $SU(2)\,$. 
The group $SU(2)$ is also the universal covering group of $SO(3)\,$,
but beware that the universal cover of
$SO(2,1)^\uparrow$ is actually ${\mathbb R}^3\,$, covering
$SO(2,1)^\uparrow$ infinitely many times. Thus one may not speak of
the spin group for the case of the proper
orthochronous Lorentz group in spacetime dimension three. The
analogue is that the universal cover of $SO(2)\cong U(1)$ is
$\mathbb R\,$, that covers  $U(1)$ infinitely many times, so that one 
may not speak of the spin group for the degenerate case of the rotation
group in two spatial dimensions.

\subsection{The Poincar\'e group and algebra}
%
The transformations
$$x^{\prime\mu}=\Lambda^{\mu}_{~\nu}x^{\mu}+a^{\mu} $$
where $a$ is
a spacetime translation vector, form the group of all
inhomogeneous Lorentz transformations. If one denotes a general
transformation by $(\Lambda,a)\,$, the multiplication law in the
Poincar\'e group is given by $$ (\Lambda_2,a_2)\cdot
(\Lambda_1,a_1)=(\Lambda_2\Lambda_1,a_2+\Lambda_2a_1)\,,$$ so that
the \textsl{inhomogeneous Lorentz group} is the semi-direct
product denoted by
$$IO(D-1,1)={\mathbb R}^D\rtimes O(D-1,1)\,.$$
The subgroup $ISO(D-1,1)^\uparrow$ of inhomogeneous proper
orthochronous Lorentz transformations is called the
\textsl{Poincar\'e group}. The Lie algebra $\mathfrak{iso}(D-1,1)$
of the Poincar\'e group is presented by the generators
$\{\,P_\mu\,,\,M_{\nu\rho}\,\}$ and by the commutation relations
\begin{eqnarray}
i\,[M_{\mu\nu},M_{\rho\sigma}] &=&
\eta_{\nu\rho}M_{\mu\sigma}-\eta_{\mu\rho}M_{\nu\sigma}-\eta_{\sigma\mu}M_{\rho\nu}
+\eta_{\sigma\nu}M_{\rho\mu}\,
\label{BeBo:1.5.10a} \\
i\,{[}P_{\mu},M_{\rho\sigma}{]} &=&
\eta_{\mu\rho}P_{\sigma}-\eta_{\mu\sigma}P_{\rho}\,,
\label{BeBo:1.5.10b} \\
i\,{[}P_{\mu},P_{\rho}{]} &=& 0\,. \label{BeBo:1.5.10c}
\end{eqnarray}
Two subalgebras must be distinguished: the Lie algebra
$\mathfrak{so}(D-1,1)$ of the Lorentz group presented by the
generators $\{\,M_{\nu\rho}\,\}$ and by the commutation relations
(\ref{BeBo:1.5.10a}), and the Lie algebra ${\mathbb R}^D$ of the
Abelian translation group presented by the generators
$\{\,P_\mu\,\}$ and by the commutation relations
(\ref{BeBo:1.5.10c}). The latter algebra ${\mathbb R}^D$ is an
ideal of the Poincar\'e algebra, as can be seen from
(\ref{BeBo:1.5.10b}). Altogether, this implies that the Lie
algebra of the Poincar\'e group is the semi-direct sum
$\mathfrak{iso}(D-1,1)={\mathbb
R}^D\niplus\mathfrak{so}(D-1,1)\,$.

The Casimir elements of a Lie algebra $\mathfrak{g}$ are
homogeneous polynomials in the generators of $\mathfrak{g}$
providing a distinguished basis of the center ${\cal Z}\Big({\cal
U}(\mathfrak{g})\Big)$ of the universal enveloping algebra ${\cal
U}(\mathfrak{g})$ (see \textit{e.g.} the part V of the lecture
notes \cite{BeBo:Liealgs}). The quadratic Casimir operator of the
Lorentz algebra $\mathfrak{so}(D-1,1)$ is the square of the
generators $M_{\mu\nu}$: 
\begin{equation}{\cal C}_2\Big(\mathfrak{so}(D-1,1)\Big) \,=\,
\frac12\,M^{\mu\nu}M_{\mu\nu}\,.
\label{BeBo:QuadratiC}
\end{equation} The quadratic Casimir
operator of the Poincar\'e algebra $\mathfrak{iso}(D-1,1)$ is the
square of the momentum
\begin{equation}{\cal C}_2\Big(\mathfrak{iso}(D-1,1)\Big) \,=\,
-P^{\mu}P_{\mu}\,, \label{BeBo:quadratiC}\end{equation} while the
quartic Casimir operator is
\begin{equation}
{\cal C}_4\Big(\mathfrak{iso}(D-1,1)\Big) \,=\, -{1\over 2}P^2M_{\mu\nu}M^{\mu\nu}
+M_{\mu\rho}P^\rho M^{\mu\sigma}P_\sigma\,,
\label{BeBo:QuartiC}
\end{equation}
which, for $D=4$,
is the square of the Pauli-Lubanski vector $W^{\mu}$, 
$$W^{\mu}:=\frac{1}{2}\,\varepsilon^{\mu\nu\rho\sigma}M_{\nu\rho}P_{\sigma}\,.$$

\subsection{ABC of unitary representations}
\label{BeBo:ABC}

The mathematical property that all non-trivial unitary reps of a
non-compact simple Lie group must be infinite-dimensional has some
physical significance, as will be reviewed later.

\vspace{2mm}\noindent \textbf{Finite-dimensional unitary reps of
non-compact simple Lie groups:} \textit{Let
$\textsc{U}:G\rightarrow U(n)$ be a unitary representation of a
Lie group $G$ acting on a (real or complex) Hilbert space $\cal H$
of finite dimension $n\in\mathbb N$. Then $\textsc{U}$ is
completely reducible. Moreover, if $\textsc{U}$ is irreducible and
if $G$ is a connected simple non-compact Lie group, then
$\textsc{U}$ is the trivial representation.}

\vspace{2mm}\noindent\underline{Proof:} For the property that
$\textsc{U}$ is completely reducible, we refer \textit{e.g.} to
the proof of the proposition 5.15 in \cite{BeBo:BHall}. The image
$\textsc{U}(G)$ for any unitary representation $\textsc{U}$ defines
a subgroup of $U(n)\,$. Moreover, the kernel of $\textsc{U}$ is a
normal subgroup of the simple group $G$. Therefore, either the
representation is trivial and $\ker \textsc{U}=G\,$, or it is
faithfull and $\ker \textsc{U}=\{e\}\,$. In the latter case,
$\textsc{U}$ is invertible and its image is isomorphic to its
domain, $\textsc{U}(G)\cong G$. Actually, the image
$\textsc{U}(G)$ is a non-compact subgroup of $U(n)$ because if
$\textsc{U}(G)$ was compact, then
$\textsc{U}^{-1}\big(\textsc{U}(G)\big)=G$ would be compact since
$\textsc{U}^{-1}$ is a continuous map. But the group $U(n)$
is compact, thus it cannot contain a non-compact subgroup.
Therefore the representation cannot be faithful, so that it is
trivial. (A different proof of the second part of the theorem may
be found in the section 8.1.B of \cite{BeBo:Barut}.) \qed

\vspace{2mm} Another mathematical result which is of physical
significance is the following theorem on unitary irreducible
representations (UIRs) of compact Lie groups.

\vspace{2mm}\noindent \textbf{Unitary reps of compact Lie groups:}
\textit{Let $\textsc{U}$ be a UIR of a compact Lie group $G$,
acting on a (real or complex) Hilbert space $\cal H$. Then $\cal
H$ is finite-dimensional. Moreover, every unitary representation
of $G$ is a direct sum of UIRs (the sum may be infinite).}

\vspace{2mm}\noindent\underline{Proof:} The proofs are somewhat
lengthy and technical so we refer to the section 7.1 of
\cite{BeBo:Barut} for complete details. \qed

\vspace{2mm}\noindent\textbf{Examples of (not so) simple groups:}

\noindent $\bullet$ On the one hand, all (pseudo)-orthogonal
groups $SO(p,q)$ are either Abelian ($p+q= 2$), 
non-simple ($p+q = 4$) or simple ($p+q=3$ and $p+q>4\,$).
Moreover, the orthogonal groups ($p\,q=0$) are compact, while
the pseudo-orthogonal groups ($p\,q\neq0$) are non-compact.

\noindent $\bullet$ On the other hand, the inhomogeneous Lorentz
group $IO(D-1,1)$ is non-compact (both ${\mathbb R}^D$ and
$O(D-1,1)$ are non-compact) but it is \textit{not} semi-simple
(because its normal subgroup ${\mathbb R}^D$ is Abelian).

\section{Elementary particles as unitary irreducible representations
of the isometry group} \label{BeBo:element}

Except for the final remarks, this section is based almost \textit{ad
verbatim} on the introduction of the illuminating work of Bargmann
and Wigner \cite{BeBo:Bargmann48}, modulo some changes of notation
and terminology in order to follow the modern conventions.

The wave functions $\mid\psi\,\,\rangle$ describing the possible
states of a quantum-mechanical system form a linear vector space
$\cal H$ which, in general, is infinite-dimensional and on which a
positive-definite inner product
$\langle\,\,\phi\mid\psi\,\,\rangle$ is defined for any two wave
functions $\mid\phi\,\,\rangle$ and $\mid\psi\,\,\rangle$
(\textit{i.e.} they form a Hilbert space). The inner product
usually involves an integration over the whole configuration or
momentum space and, for particles of non-vanishing spin, a
summation over the spin indices.

If the wave functions in question refer to a free particle and
satisfy relativistic wave equations, there exists a correspondence
between the wave functions describing the same state in different
Lorentz frames. The transformations considered here form the group
of all \textit{inhomogeneous} Lorentz transformations (including
translations of the origin in space and time). Let
$\mid\psi\,\,\rangle$ and $\mid\psi\,\,\rangle^\prime$ be the wave
functions of the same state in two Lorentz frames $L$ and
$L^\prime$, respectively. Then
$\mid\psi\,\,\rangle^\prime=U(\Lambda,a)\mid\psi\,\,\rangle$,
where $U(\Lambda,a)$ is a linear unitary operator which depends on
the transformation $(\Lambda,a)$ leading from $L$ to $L^\prime\,$.
By a proper normalization, $U$ is determined by $\Lambda$ up to a
factor $\pm 1\,$. Moreover, the operators $U$ form a single- or
double-valued representation of the inhomogeneous Lorentz group,
\textit{i.e.}, for a succession of two transformations
$(\Lambda_1,a_1)$ and $(\Lambda_2,a_2)$, we have
\begin{equation}
U(\Lambda_2,a_2)U(\Lambda_1,a_1)=\pm
U(\Lambda_2\Lambda_1,a_2+\Lambda_2a_1)\,.
\label{BeBo:compositionrule}
\end{equation}

Since all Lorentz frames are equivalent for the description of our
system, it follows that, together with $\mid\psi\,\,\rangle\,$,
$U(\Lambda,a)\mid\psi\,\,\rangle$ is also a possible state viewed
from the original Lorentz frame $L\,$. Thus, the vector space
$\cal H$ contains, with every $\mid\psi\,\,\rangle\,$, all
transforms $U(\Lambda,a)\mid\psi\,\,\rangle\,$, where
$(\Lambda,a)$ is any inhomogenous Lorentz transformation.

The operators $U$ may also replace the wave equation of the
system. In our discussion, we use the wave functions in the
``Heisenberg" representation, so that a given
$\mid\psi\,\,\rangle$ represents the system for all times, and may
be chosen as the ``Schr\"{o}dinger" wave function at time $t=0$
in a given Lorentz frame $L$. To find
$\mid\psi\,\,\rangle_{t_0}\,$, the Schr\"{o}dinger function at
time $t_0\,$, one must therefore transform to a frame $L^\prime$
for which $t^\prime=t-t_0\,$, while all other coordinates remain
unchanged. Then
$\mid\psi\,\,\rangle_{t_0}=U(\Lambda,a)\mid\psi\,\,\rangle\,$,
where $(\Lambda,a)$ is the transformation leading from $L$ to
$L^\prime\,$.

A classification of all unitary representations of the
inhomogeneous Lorentz group, \textit{i.e.} of all solution of
(\ref{BeBo:compositionrule}), amounts, therefore, to a
classification of all possible relativistic wave equations. Two
reps $U$ and $\widetilde{U}=VUV^{-1}\,$, where $V$ is a fixed
unitary operator, are equivalent. If the system is described by
wave functions $\mid\psi\,\,\rangle\,$, the description by
\begin{equation}
\widetilde{\mid\psi\,\,\rangle}=V\mid\psi\,\,\rangle
\label{BeBo:equivalentrep}
\end{equation}
is isomorphic with respect to linear superposition, with respect
to forming the
inner product of two wave functions, and also with respect to the transition from
one Lorentz frame to another. In fact, if
$\mid\psi\,\,\rangle^\prime=U(\Lambda,a)\mid\psi\,\,\rangle\,$,
then
$$\widetilde{\mid\psi\,\,\rangle}^\prime=V\mid\psi\,\,\rangle^\prime
=VU(\Lambda,a)V^{-1}\widetilde{\mid\psi\,\,\rangle}=
\widetilde{U}(\Lambda,a)\widetilde{\mid\psi\,\,\rangle}\,.$$ Thus,
one obtains classes of equivalent wave equations. Finally, it is
sufficient to determine the irreducible representations (irreps)
since any other may be built from them.

Two descriptions which are equivalent according to
(\ref{BeBo:equivalentrep}) may be quite different in appearance.
The best known example is the description of the electromagnetic
field by the field strength and the vector potential, respectively.
It cannot be claimed either that equivalence in the sense of
(\ref{BeBo:equivalentrep}) implies equivalence in every physical
aspect. 
It should be emphasized
that any selection of one among the equivalent systems
involves an explicit or implicit assumption as to possible
interactions, \textit{etc}. Our analysis is necessarily restricted
to free particles and does not lead to any assertion about
possible interactions.

The present discussion is not based on any hypothesis about the
structure of the wave equations provided that they be covariant.
In particular, it is not necessary to assume differential
equations in configuration space. But it is a result of the
group-theoretical analysis that every irreducible field equation is
equivalent, in the sense of (\ref{BeBo:equivalentrep}), to a
system of differential equations for fields on Minkowski
spacetime.

\newpage\noindent\textbf{Remarks:}

\noindent $\bullet$ An important theorem proved by Wigner is that
any symmetry transformation that is continuously related to the
identity must be represented by a linear unitary operator (see
\textit{e.g.} the appendix A of \cite{BeBo:Weinberg}). Strictly
speaking, physical states are represented by \textit{rays} in a
Hilbert space. Therefore the unitary representations of the
symmetry group are actually understood to be \textit{projective}
representations. In spacetime dimensions $D\geqslant 4\,$, 
this subtlety\footnote{The case  $D=3$ is even more subtle 
and is  treated in Appendix \ref{App:A}.} reduces to the standard 
distinction between single and double valued representations of the 
Poincar\'e group, as was taken for granted in the text.

\noindent $\bullet$ Notice that the previous discussion remains
entirely valid if the Minkowski spacetime ${\mathbb R}^{D-1,1}$ is
replaced everywhere by any other maximally symmetric spacetime
(\textit{i.e.} de Sitter spacetime $dS_D$, or anti de Sitter
spacetime $AdS_D$) under the condition that the inhomogeneous
Lorentz group $IO(D-1,1)$ be also replaced everywhere by the
corresponding group of isometries (respectively, $\,O(D,\,1)$ or,
$O(D-1,\,2)\,$).

\noindent $\bullet$ In first-quantization, particles are described
by fields on the spacetime and isometries are represented by
unitary operators. A particle is said to be ``elementary" if the
representation is irreducible, and ``composite" if the
representation is made of a product of irreps. 

\noindent $\bullet$  A modern point of view on Quantum Field 
Theory \cite{BeBo:Weinberg} is that a quantum field (not to be 
confused with the state vector discussed above) 
is an \emph{operator} defined at each point of space and time, 
that acts in a Fock space of states, the field being represented 
by a superposition, for different values of the momentum, 
of one-particule annihilation and creation operators for particle 
and the associated antiparticle. 
The approach of \cite{BeBo:Weinberg} is to build up the 
quantum field by imposing Lorentz invariance at every stage.
To quote Weinberg, the field equation satisfied by 
the \emph{quantum} field arises almost incidentally, as a 
byproduct of his construction. 

\noindent $\bullet$ A unitary representation 
of the isometry group describes the one-particle Hilbert space of states.
The group-theoretical argument of Bargmann 
and Wigner \cite{BeBo:Bargmann48} applies to the one-particule 
states of a free particle.\footnote{See e.g. Eq. (2.5.1) of \cite{BeBo:Weinberg}
where the one-particle state vectors are denoted by $\Psi_{p,\sigma}\,$.} 
The classification of the UIRs of the Poincar\'e group indeed yields 
the Klein-Gordon equation for a massive particle, or the D'Alembert equation 
in the case of a massless particle \cite{BeBo:Bargmann48}. 
This comes automatically from the group-theoretical analysis and is 
\emph{not} an assumption. 

\vspace{2mm}\noindent\textbf{Summary:} On the one
hand, the rules of quantum mechanics imply that quantum symmetries
correspond to unitary representations of the symmetry group
carried by the Hilbert space of physical states. Furthermore, if
time translations constitute a one-parameter subgroup of the symmetry
group, then the Schr\"{o}dinger equation for the time evolution of a 
state vector essentially is a unitary representation of this subgroup. 
On the other hand, the principle of relativity dictates that the isometries 
of spacetime be symmetries of the physical system. 
All together, this implies that linear relativistic field equations may 
be identified with unitary reps of the isometry group.

\section{Classification of the unitary representations}

\subsection{Induced representations}
\label{BeBo:inducedreps}
%

The method of induced reps was introduced by Wigner in his seminal
paper \cite{BeBo:Wigner39} on the unitary representations of the
inhomogeneous Lorentz group $IO(3,1)$ in four spacetime
dimensions, which admits a straightforward generalization to any
spacetime dimension $D$, as reviewed now. The content of this
subsection finds its origin in the section 2.5 of the
comprehensive textbook \cite{BeBo:Weinberg}.

{}From (\ref{BeBo:1.5.10c}) one sees that all the translation
generators commute with each other, so it is natural to express
physical states $\mid\psi\,\,\rangle$ in terms of eigenvectors of
the translation generators $P^{\mu}\,$. Introducing a label
$\sigma$ to denote all other degrees of freedom, one thus
considers states $\Psi_{q,\sigma}$ with ${P}_{\mu}
\Psi_{q,\sigma}=q_{\mu}\Psi_{q,\sigma}\,$. From the infinitesimal
translation $U=\mathun -iP^{\mu}\epsilon_{\mu}$ and repeated
applications of it, one finds that finite translations are
represented on $\cal H$ by $U(\mathun,
a)=\exp(-i\,P^{\mu}a_{\mu})\,$, so one has $$ U(\mathun
,a)\,\Psi_{q,\sigma}\, =\, e^{-i\,q\cdot a}\,\Psi_{q,\sigma}\,. $$
Using (\ref{BeBo:1.5.10b}), one sees that the effect of operating
on $\Psi_{p,\sigma}$ with a quantum homogeneous transformation
$U(\Lambda,0)\equiv U(\Lambda)$ is to produce an eigenvector of
the translation generators with eigenvalue $\Lambda p$ :
\begin{eqnarray}
P^{\mu}U(\Lambda)\Psi_{p,\sigma}&=&U(\Lambda)[U^{-1}(\Lambda)P^{\mu}
U(\Lambda)]\Psi_{p,\sigma}=
U(\Lambda)({(\Lambda^{-1})}{}_{\rho}{}^{\mu}P^{\rho})\Psi_{p,\sigma}
\nonumber \\
&=&\Lambda^{\mu}_{~\rho}\,p^{\rho}\,U(\Lambda)\Psi_{p,\sigma}\,,\nonumber
\end{eqnarray}
since $(\Lambda^{-1}){}_{\rho}{}^{\mu} = \Lambda^{\mu}{}_{\rho}\,$.
Hence $U(\Lambda)\Psi_{p,\sigma}$ must be a linear combination of
the states $\Psi_{\Lambda p,\sigma}$ :
\begin{equation}
U(\Lambda)\Psi_{p,\sigma}=\sum_{\sigma'}C_{\sigma'\sigma}(\Lambda,p)\Psi_{\Lambda
p,\sigma'}\,. \label{BeBo:2.5.3} \end{equation} In general, it is
possible by using suitable linear combinations of the
$\Psi_{p,\sigma}$ to choose the $\sigma$ labels in such a way that
the matrix $C_{\sigma'\sigma}(\Lambda,p)$ is block-diagonal; in
other words, so that the $\Psi_{p,\sigma}$ with $\sigma$ within
any one block {\textsl{by themselves}} furnish a representation of
the Poincar\'e group. It is natural to identify the states of a
specific particle type with the components of a representation of
the Poincar\'e group which is irreducible, in the sense that it
cannot be further decomposed in this way. It is clear from
(\ref{BeBo:2.5.3}) that all states $\Psi_{p,\sigma}$ in an irrep
of the Poincar\'e group have momenta $p^{\mu}$ belonging to the orbit of
a single fixed momentum, say $q^\mu$.

One has to work out the structure of the coefficients
$C_{\sigma'\sigma}(\Lambda,p)$ in irreducible representations of
the Poincar\'e group. In order to do that, note that the only
functions of $p^{\mu}$ that are left invariant by all
transformations $\Lambda^{\mu}_{~\nu}\in SO(D-1,1)^{\uparrow}$
are, of course, $p^2=\eta_{\mu\nu}p^{\mu}p^{\nu}$ and, for
$p^2\leqslant 0\,$, also the sign of $p^0\,$. Hence, for each
value of $p^2\,$, and (for $p^2\leqslant 0$) each sign of $p^0\,$,
one can choose a standard four-momentum, say $q^{\mu}\,$, and
express any $p^{\mu}$ of this class as $$
p^{\mu}=L^{\mu}_{~\nu}(p)q^{\nu}\,, $$ where $L^{\mu}_{~\nu}$ is
some standard proper orthochronous Lorentz transformation that
depends on $p^{\mu}\,$, and also implicitly on our choice of
$q^{\mu}\,$. One can define the states $\Psi_{p,\sigma}$ of
momentum $p^{\mu}$ by \begin{equation} \Psi_{p,\sigma}\equiv
N(p)\,U\Big(L(p)\Big)\,\Psi_{q,\sigma}\,, \label{BeBo:2.5.5}
\end{equation} where $N(p)$ is a numerical normalization factor.
Operating on (\ref{BeBo:2.5.5}) with an arbitrary homogeneous
Lorentz transformation $U(\Lambda)\,$, one now finds
\begin{eqnarray} U(\Lambda)\Psi_{p,\sigma}&=&N(p)\,U\Big(\Lambda L(p)\Big)\,
\Psi_{q,\sigma}\nonumber \\
&=& N(p)\,U\Big(L(\Lambda p)\Big)\,U\Big(L^{-1}(\Lambda p)\Lambda
L(p)\Big)\Psi_{q,\sigma}\,. \label{BeBo:2.5.6} \end{eqnarray} The
point of this last step is that the Lorentz transformation
$L^{-1}(\Lambda p)\Lambda L(p)$ takes $q$ to $L(p)q=p\,$, then to
$\Lambda p\,$, and finally back to $q\,$, so it belongs to the
subgroup of the Lorentz group consisting of Lorentz
transformations $W^{\mu}_{~\nu}$ that leave $q^{\mu}$ {invariant
:} $W^{\mu}_{~\nu}q^{\nu}=q^{\mu}\,$. This stability subgroup is
called the \textsl{little group} corresponding to $q\,$. For any $W,\bar{W}$ in the
little group, one has \begin{equation}
U(W)\Psi_{q,\sigma}=\sum_{\sigma'}D^q_{\sigma'\sigma}(W)\Psi_{q,\sigma'}
\label{BeBo:2.5.8} \end{equation} and
$$D^q_{\sigma'\sigma}(\bar{W}W)=\sum_{\sigma''}D^q_{\sigma'\sigma''}(\bar{W})D^q_{\sigma''\sigma}(W)\,,$$
that is to say, the coefficients $D^q(W)$ furnish a representation
of the little group. In particular, for 
$W(\Lambda,p)\equiv L^{-1}(\Lambda p)\Lambda L(p)\,$, 
the equation (\ref{BeBo:2.5.6}) becomes $$
U(\Lambda)\Psi_{p,\sigma}=N(p)\sum_{\sigma'}D_{\sigma'\sigma}(W(\Lambda,p))U\Big(L(\Lambda
p)\Big)\Psi_{q,\sigma'} $$ or, recalling the definition
(\ref{BeBo:2.5.5}), \begin{equation}
U(\Lambda)\Psi_{p,\sigma}=\frac{N(p)}{N(\Lambda
p)}\sum_{\sigma'}D_{\sigma'\sigma}\big(W(\Lambda,p)\big)\Psi_{\Lambda
p,\sigma'}\,. \label{BeBo:2.5.11} \end{equation} Apart from the
question of normalization, the problem of determining the
coefficients $C_{\sigma'\sigma}$ in the transformation rule
(\ref{BeBo:2.5.3}) has been reduced to the problem of determining
the coefficients $D_{\sigma'\sigma}$. In other words, the problem
of determining all possible irreps of the Poincar\'e group has
been reduced to the problem of finding all possible irreps of the
little group, depending on the class of momentum to which
$q^{\mu}$ belongs. This approach, of deriving representations of a
semi-direct product like the inhomogeneous Lorentz group from the
representations of the stability subgroup, is called the
\textsl{method of induced representations}.

The wave function $\Psi_{p,\sigma}$ depends on the momentum,
therefore its Fourier transform $\Psi_{x,\sigma}$ depends on the
spacetime coordinate, so that the wave function is called a
(complex) \textsl{field} on Minkowski spacetime ${\mathbb
R}^{D-1,1}$ and the quantities $\Psi_{x,\sigma}$ at fixed $x$ and for
varying $\sigma$ are referred to as its \textsl{physical components}
at $x\,$.

\subsection{Orbits and stability subgroups}
\label{BeBo:orbit}
%

The orbit of a given non-vanishing vector $q^\mu$ of Minkowski
spacetime ${\mathbb R}^{D-1,1}$ under Lorentz transformations is,
by definition, the hypersurface of constant momentum square
$p^2\,$. Geometrically speaking, it is a quadric of curvature
radius $m>0$. More precisely, the hypersurface
\begin{itemize}
  \item $p^2=-m^2$ is a two-sheeted hyperboloid, each sheet of which is
called a \textsl{mass-shell}. The corresponding UIR is said to be
\textsl{massive}.
  \item $p^2=0$ is a cone, each half of which is called a
  \textsl{light-cone}. The corresponding UIR is said to
be \textsl{massless} ($m=0$).
  \item $p^2=+m^2$ is a one-sheeted hyperboloid. The corresponding UIR is said to
be \textsl{tachyonic}.
\end{itemize}
Orthochronous Lorentz transformations preserve the sign of the
time component of vectors of non-positive momentum-squared, thus
the orbit of a time-like (light-like) vector is the mass-shell
(respectively, light-cone) to which the corresponding vector belongs.

\vspace{2mm}\noindent\textbf{Remarks:}

\noindent $\bullet$ Notice that the Hilbert space carrying the
irrep is indeed an eigenspace of the quadratic Casimir operator
(\ref{BeBo:quadratiC}),
the eigenvalue of which is ${\cal C}_2=\pm\, m^2\,$ (the
eigenvalue is real because the representation is unitary), 
as it should according to Schur's lemma.
Moreover, the quadratic Casimir classifies the UIRs but does not
entirely characterize them.

{\noindent $\bullet$
For any pair of fields that transform in the same UIR of the Poincar\'e group, 
the Poincar\'e-invariant scalar product for $p^{2}\leqslant 0$ is given by
$\langle \Psi \vert \Phi\rangle = \int d^{D}p\, \delta(p^{2}+m^{2})\Theta(p^{0})
\sum_{\sigma} \Psi^{*}_{p,\sigma}\Phi_{p,\sigma}\,$.}

\noindent $\bullet$ As quoted in Section \ref{BeBo:element}, it is
not necessary to assume differential equations in position space,
because the group-theoretical analysis directly leads to a wave
function which is a function of the momenta on the orbit, the
Fourier transform of which is a function in position space obeying
the Klein--Gordon equation $\Box\,\Psi_{x,\sigma}\,=\,\pm\,
m^2\,\Psi_{x,\sigma}\,$. By a slight abuse of terminology, states or fields that satisfy their relativistic equations of motion
are called ``on-(mass-)shell" in physics literature, 
while those for which those equations have not
been imposed are called ``off-shell".
\vspace{2mm}

By going to a rest-frame, it is easy to show that the stabilizer
of a time-like vector $q^\mu=(m,\overrightarrow{0})\neq 0$ is the
rotation subgroup $SO(D-1)\subset SO(D-1,1)^\uparrow$. For a
space-like vector, one may choose a frame such that the
non-vanishing momentum is along the $(D-1)$th spatial axis:
$q^\mu=(0,0,\ldots,0,m)\neq 0$. Thus its stabilizer is the
subgroup $SO(D-2,1)^\uparrow\subset SO(D-1,1)^\uparrow$. In the
case of a light-like vector, the little group ``\textit{is not
quite so obvious}" to determine, as was stressed by Wigner himself
\cite{BeBo:Wigner63}. It clearly contains the rotation group
$SO(D-2)$ in the space-like hyperplane ${\mathbb R}^{D-2}$
transverse to the light-ray along the momentum. Now, we will
provide an algebraic proof that the stabilizer of a light-like
vector is the Euclidean group $ISO(D-2)\,$. According to Wigner,
reviewing his $D=4$ analysis, ``\textit{no simple argument is
known (...) to show directly that the group of Lorentz
transformations which leave a null vector invariant is isomorphic
to the two-dimensional Euclidean group, desirable as it would be
to have such an argument. Clearly, there is no plane in the
four-space of momenta in which these transformations could be
interpreted directly as displacements (...) because all
transformations considered here are homogeneous}"
\cite{BeBo:Wigner63}. Even though there is no simple geometric way
to understand this fact, the algebraic proof reviewed here is
rather straightforward.

\vspace{2mm}\noindent\underline{Proof:} By going in a light-cone
frame (see Section \ref{BeBo:Lorentzgr}), it is possible to
express the components of a momentum $p^\mu$ obeying $p^2=0$ as
$p_{\mu}=(p_{-},0,0,\ldots, 0)\,$. In words, one can set the
component $p_+$ to zero, as well as all the transverse components
$p_m\,$ ($m=1,\ldots,D-2$). The condition that the component $p_-$
be unaffected by a Lorentz transformation translates as
$0\stackrel{!}{=}i[p_-,M_{\nu\rho}]=\eta_{-\nu}\,p_{\rho}-\eta_{-\rho}\,p_{\nu}$
due to (\ref{BeBo:1.5.10b}). Obviously, the transformation
generated by $M_{+-}$ does modify $p_-$, hence it cannot be part
of the little group for $p\,$. The other Lorentz generators
preserve $p_-\,$, but they should also satisfy the equations
$[p_m,M_{\mu\nu}]=0=[p_+,M_{\mu\nu}]\,$. It is readily seen that
$i[p_m,M_{n-}]=\delta_{mn}p_-\ne 0$ (for $m=n$), therefore
$M_{n-}$ does not belong to the little group of $p_\mu$ either. We
are left with the generators $\{M_{mn}, M_{+n}\}$ which preserve
the (vanishing) value of $p_{+}\,$. It turns out to be more
convenient for later purpose to work with the generators
$\pi_n:=p_-M_{+n}=p^\mu M_{\mu\, n}$ instead. This redefinition
does not modify the algebra since $p_-$ commutes with all the
generators of the little group. {}From the Poincar\'e algebra
(\ref{BeBo:1.5.10a})--(\ref{BeBo:1.5.10c}) one finds, in the
light-cone frame,
\begin{eqnarray}
i\,[M_{mn},M_{pq}] &=&
\delta_{np}M_{mq}-\delta_{mp}M_{nq}-\delta_{qm}M_{pn}
+\delta_{qn}M_{pm}\,,
\label{BeBo:1.5.10lca} \\
i\,[\pi_m,M_{np}] &=& \delta_{mn}\pi_{p}-\delta_{mp}\pi_{n}\,,
\label{BeBo:1.5.10lcb} \\
i\,{[}\pi_m,\pi_n{]} &=& 0\,. \label{BeBo:1.5.10cc}
\end{eqnarray}
As can be seen, the generators $\{M_{mn},\pi_m\}$ span the Lie
algebra of the inhomogeneous orthogonal group $ISO(D-2)\,$. \qed

For later purpose, notice that the quadratic Casimir operator of
the Euclidean algebra $\mathfrak{iso}(D-2)$ presented by the
generators $\{M_{mn},\pi_m\}$ and the relations
(\ref{BeBo:1.5.10lca})-(\ref{BeBo:1.5.10cc}) is the square of the
``translation" generators
\begin{equation}{\cal C}_2\Big(\mathfrak{iso}(D-2)\Big) \,=\,
\pi^m\pi_m\,. \label{BeBo:quadratinfinitespin}
\end{equation}

To end up this discussion, one should consider the case of a
vanishing momentum. Of course, the orbit of a vanishing vector
under linear transformations is itself while its stabilizer is the
whole linear subgroup. Therefore, the subgroup of
$SO(D-1,1)^{\uparrow}$ leaving invariant the zero-momentum vector
$p^\mu=0$ is the whole group itself. This ends up the
determination of the orbit and stabilizer of any possible vector
$\in {\mathbb R}^{D-1,1}\,$.

\vspace{2mm}\noindent\textbf{Remark:} The zero-momentum
($q^\mu=0$) representations are essentially UIRs of the little
group $SO(D-1,1)^{\uparrow}$ because the translation group acts
trivially. The proper orthochronous Lorentz group may be
identified with the isometry group of the de Sitter spacetime
$dS_{D-1}$. In other words, the wave function of the zero-momentum
representation may be interpreted as a wave function on a
lower-dimensional de Sitter spacetime, and conversely. Even though
their physical meaning may differ, both UIRs may be mathematically
identified.\vspace{1mm}

\subsection{Classification}
\label{BeBo:classification}
%

To summarise the previous subsection, the UIRs of the Poincar\'e
group $ISO(D-1,1)^\uparrow$ have been divided into four classes
according to the four possible orbits of the momentum, 
summarised in the following
table (where $m^2>0\,$):
\begin{center}
\begin{tabular}{|c|c|c|c|}
 \hline
  Gender & Orbit & Stability subgroup & UIR \\\hline\hline
  $p^2=-m^2$ & Mass-shell & $SO(D-1)$ & \small{Massive} \\\hline
  $p^2=0$ & Light-cone & $ISO(D-2)$ & \small{Massless} \\\hline
  $p^2=+m^2$ & Hyperboloid & $SO(D-2,1)^{\uparrow}$  & \small{Tachyonic} \\\hline
  $p_\mu=0$ & Origin & $SO(D-1,1)^{\uparrow}$ & \small{Zero-momentum} \\ \hline
\end{tabular}
\end{center}
\vspace{1mm}The problem of classifying the UIRs of the Poincar\'e
group $ISO(D-1,1)^\uparrow$ has been reduced to the classication
of the UIRs of the stability subgroup of the momentum, which are
either a unimodular orthogonal group, an Euclidean group or a
proper orthochronous Lorentz group.

Actually, the method of induced representation may also be applied
to the classification of the UIRs of the Euclidean group
$ISO(D-2)$, the little group of a massless particle. The important
thing to understand is that the light-like momentum $p^\mu$ is
fixed and that what should be examined is the action of the little
group on the physical components characterized by $\sigma\,$. From
(\ref{BeBo:1.5.10cc}) one sees that the $D-2$ ``translation"
generators $\pi^i$ commute with each other, so it is natural to
express physical states $\Psi_{p,\sigma}$ in terms of eigenvectors
$\xi^m$ of these generators $\pi^m$. Introducing a label
$\varsigma$ to denote all remaining physical components, one thus
considers states $\Psi_{p,\,\xi,\,\varsigma}$ with $\pi_m
\Psi_{p,\,\xi,\,\varsigma}=\xi_m\Psi_{p,\,\xi,\,\varsigma}\,$. The
discussion presented in Subsection \ref{BeBo:inducedreps} may be
repeated almost identically, up to the replacement of the momentum
$p$ by the eigenvector $\xi$, the label $\sigma$ by
$\varsigma$, the Poincar\'e group $ISO(D-1,1)^\uparrow$ by the
Euclidean group $ISO(D-2)$ and the proper orthochronous Lorentz
group $SO(D-1,1)^\uparrow$ by the unimodular orthogonal group
$SO(D-2)\,$. The conclusion is therefore similar: the problem of
determining all possible irreps of the massless little group
$ISO(D-2)$ has been reduced to the problem of finding all possible
irreps of the stability subgroup of the $(D-2)$-vector $\xi\,$,
called the \textsl{short little group} in the literature
\cite{BeBo:Brink2002}.

The massless representations induced by a non-trivial
representation of the little group may therefore be divided into
distinct categories, depending on the class of momentum to which
$\xi^m$ belongs. The situation is simpler here because there exist
only two possible classes of orbits for a vector in the Euclidean
space ${\mathbb R}^{D-2}$: either the origin $\xi^m=0\,$, or a
$(D-3)$-sphere of radius $\mu>0\,$. In the first case, the action
of the elusive ``translation" operators $\pi^m$ is trivial and,
effectively, the little group is identified with the short little
group $SO(D-2)$. These representations are most often referred to
as \textsl{helicity} representations by analogy with the $D=4$
case. 
In the second case, the corresponding representations are
most often referred to as \textsl{continuous spin} representations
\cite{BeBo:Brink2002}, even though Wigner also used the name
\textsl{infinite spin} \cite{BeBo:Wigner63}. The former name
originates from the fact that the transverse vector $\xi^m$ has a
continuous range of values. Nevertheless, the latter name is more
adequate in some respect, as will be argued later on. Roughly
speaking the point is that, on the orbit $\xi^2=\mu^2$, the
components spanned by the internal vector $\xi^m$ take values on
the sphere $S^{D-3}\subset{\mathbb R}^{D-2}$ of radius
$\mu=|\xi|\,$. The ``radius" $\mu$ of this internal sphere has
actually the dimension of a mass parameter (the reason is that the
sphere $S^{D-3}$ is somehow in internal ``momentum" space).
Indeed, for massless representations, the parameter $\mu$
classifying the various irreps should be understood as the
analogue of the mass for massive irreps, while the angular
coordinates on the sphere $S^{D-3}$ are the genuine ``spin"
degrees of freedom, the Fourier conjugates of which are discrete
variables as is more usual for spin degrees of freedom. This point
of view motivates the name ``infinite spin."\footnote{Actually, in
Subsection \ref{BeBo:masslessrep} an explicit derivation of the
continuous spin representation from a proper ``infinite spin"
limit of a massive representation is reviewed. All the former
comments find their natural interpretation in this point of view.}

To summarise, the UIRs of the Euclidean group $ISO(D-2)$ are
divided into two classes according to the two possible orbits 
of the $(D-2)$-vector
$\xi_m$, summarised in the following table:
\begin{center}
\begin{tabular}{|c|c|c|c|}
 \hline
 Gender & Orbit & Stability subgroup & Massless UIR \\\hline\hline
  \small{$\xi^2\,=\,\mu^2$} & Sphere & $SO(D-3)$ & \small{Infinite spin} \\\hline
  \small{$\xi_m\,=\,0$} & Origin & $SO(D-2)$ & \small{Helicity} \\\hline
\end{tabular}
\end{center}
\vspace{1mm}

As a consequence of the method of induced representations, the
physical components of a first-quantized elementary particle span
a UIR of the little group. The number of local degrees of freedom
(or of physical components) of the field theory is thus given by
the dimension of the Hilbert space carrying the UIR of the little
group. In the light of the standard results of representation
theory (reviewed in Subsection \ref{BeBo:ABC}) and using the method
of induced representation, the UIRs of the Poincar\'e group
may alternatively be divided into two distinct classes: the
\textsl{finite-component} ones (the massive and the helicity reps)
for which the (short) little group is compact, and the
\textsl{infinite-component} ones (the infinite-spin, the tachyonic
and the zero-momentum reps) for which the little group is
non-compact.

\vspace{2mm}\noindent\textbf{Remarks:}

\noindent$\bullet$ More precisely, the lower-dimensional cases
$D=2,3$ are degenerate in the following sense (when $p^\mu\neq
0$). In $D=2\,$, all little groups are trivial, thus all physical
fields are scalars. In $D=3\,$, all little groups are Abelian
(massive: $SO(2)$, massless: $\mathbb R$, tachyonic:
$SO(1,1)^\uparrow\cong\mathbb R$) hence all their UIRs have
(complex) dimension one: generically, fields have one physical degrees of
freedom. Notice
that the helicity reps may be assigned a ``\textit{conformal}
spin" if they are extended to irreps of the group $SO(D,2)\supset
SO(D-1,1)^\uparrow\,$. Notice also that the ``spin" of a massive
representation is not discretized in $D=3$ but can be an arbitrary
real number\footnote{This peculiarity is related to the existence
of anyons in three spacetime dimensions, cf. Appendix \ref{App:A}.} \cite{BeBo:Binegar} because the
universal cover of $SO(2,1)^\uparrow$ covers it infinitely often.

\noindent$\bullet$ For massive and helicity representations, the
number of local physical degrees of freedom may be determined from
the well known formulas for the dimension of any UIR of the
orthogonal groups (reviewed in Subsection \ref{BeBo:ortho} for the
tensorial irreps).

\noindent$\bullet$ This group-theoretical analysis does not probe
topological theories (such as Chern-Simons theory) because such theories
correspond to identically vanishing representations of the
little group since they have no \textit{local} physical
degrees of freedom.

\vspace{2mm}The following corollary provides a group-theoretical
explanation of the fact that combining the principle of relativity
with the rules of quantum mechanics necessarily leads to
\emph{field} theory.

\vspace{2mm}\noindent\textbf{Corollary:} \textit{Every non-trivial
unitary irreducible representation of the isometry group of any
maximally-symmetric spacetime is infinite-dimensional.}

\vspace{2mm}\noindent\underline{Proof:} The Hilbert space carrying
a non-trivial unitary representation of the Poincar\'e group is
infinite-dimensional because (i) in the generic case, $q_\mu\neq
0$, the orbit is either a hyperboloid ($p^2\neq 0$) or a cone
($p^2=0$) and the space of wave functions on the orbit is
infinite-dimensional, (ii) the zero-momentum representations of
the Poincar\'e group are unitary representations of the de Sitter
isometry group. Thus, the proof is ended by noticing that all
non-trivial unitary representations of the isometry group of
(anti) de Sitter spacetimes $(A)dS_D$ also are
infinite-dimensional, because their isometry groups are
\emph{pseudo}-orthogonal Lie groups.\qed

\section{Tensorial representations and Young diagrams}
\label{BeBo:Orthorepth}

Most of the material reviewed here may be found in textbooks such
as \cite{BeBo:Hamermesh}. Nevertheless, large parts of this
section are either copied or adapted from the
reference \cite{BeBo:Kingetal} because altogether it provides an
excellent summary, both for its pedagogical and comprehensive
values. The material collected in the present section goes slightly
beyond what is strictly necessary for these
lectures, but the reader may find it useful in specific
applications.

\subsection{Symmetric group}
\label{BeBo:permutationgr}
%

An (unlabeled) \textsl{Young diagram}, consisting of $n$ boxes
arranged in $r$ (left justified) rows, represents a
{\textsl{partition}} of the integer $n$ into $r$ {\textsl{parts}}:
\begin{eqnarray}
    n=\sum_{a=1}^r \lambda_a \qquad (\lambda_1\geqslant \lambda_2\geqslant\ldots
    \geqslant \lambda_r  )\,.
    \nonumber
\end{eqnarray}
That is, $\lambda_a$ is the number of boxes in the $a$th row.
Successive row lengths are non-increasing from top to bottom. A
simpler notation for the partition is the list of its parts:
$\lambda = \{\lambda_1,\lambda_2,\ldots,\lambda_r\}\,$. For
instance,
\begin{eqnarray}
\begin{picture}(30,30)(0,5)
\put(-30,20){$\{ 3,3,1\}=$}
\multiframe(25,25)(10.5,0){3}(10,10){}{}{}
\multiframe(25,14.5)(10.5,0){3}(10,10){}{}{}
\multiframe(25,4)(10.5,0){1}(10,10){} \put(80,20){.}
\end{picture}
\nonumber
\end{eqnarray}

\vspace{2mm}\noindent\textbf{Examples:} There are five partitions
of 4:
\begin{eqnarray}
    \{4\}, \{3,1\}, \{2,2\}, \{2,1,1\}, \{1,1,1,1\}\,.
    \label{BeBo:4.2}
\end{eqnarray}

Partitions play a key role in the study of the symmetric group
$\mathfrak{S}_n\,$. This is the group of all permutations of $n$
objects. It has $n!$ elements and {\emph{its inequivalent
irreducible representations may be labeled by the partitions of
$n\,$}}. [In the following, Greek letters $\lambda$, $\mu$ and
$\nu$ will be used to specify not only partitions and Young
diagrams but also irreducible representations of the symmetric
group and other groups.]

The connection between the symmetric group and tensors was
initially developed by H.~Weyl. This connection can be approached
in (at least) two equivalent ways.
\begin{enumerate}
    \item[\bf{A}.] Let $T_{\mu_1\ldots\,\mu_n}$ be a `generic' $n$-index tensor,
    without any special symmetry property. [For the moment, `tensor' just means
    a function of $n$ indices, not necessarily with any geometrical realization.
    It must be meaningful, however, to {\emph{add}} --- and form linear
    combinations of --- tensors of the same rank.] A {\textsl{Young tableau}}, or
    labeled Young diagram, is an assignment of the numbers $1,2,\ldots,n$ to
    the $n$
    boxes of a Young diagram $\lambda\,$.
    The tableau is {\textsl{standard}} if the numbers are
    increasing both along rows from left to right and down columns from top to
    bottom. The entries $1,\ldots,n$ in the tableau indicate the $n$ successive
    indices of $T_{\mu_1\ldots\mu_n}\,$. The tableau defines a certain
    symmetrization operation on these indices: {\emph{symmetrize}} on the set of
    indices indicated by the entries in each row, then {\emph{antisymmetrize}}
    the result on the set of indices indicated by the entries in each column.
    The resulting object is a tensor, $\widetilde{T}$, with certain index
    symmetries.
    Now let each permutation of $\mathfrak{S}_n$ act (separately) upon
    $\widetilde{T}\,$. The $n!$ results are not linearly independent; they
    span a vector space $V_{\lambda}^{\mathfrak{S}_n}$
    which supports an irreducible representation of $\mathfrak{S}_n\,$.
    Different tableaux corresponding to the same diagram $\lambda$
    yield equivalent (by not identical) representations.

    \vspace{2mm}\noindent\textbf{Example:} The partition $\{2,2\}$ of $4$ has two standard
    tableaux:
     \begin{eqnarray}
   \begin{picture}(160,20)(0,15)
   \multiframe(25,25)(10.5,0){2}(10,10){1}{2}
   \multiframe(25,14.5)(10.5,0){2}(10,10){3}{4}
   \put(65,20){and}
   \multiframe(100,25)(10.5,0){2}(10,10){1}{3}
   \multiframe(100,14.5)(10.5,0){2}(10,10){2}{4}
   \put(160,20){.}
   \end{picture}
   \label{BeBo:4.3}
   \end{eqnarray}
   Let us construct the symmetrized tensor $\widetilde{T}_{abcd}:=R_{ab|cd}$ corresponding
   to the second of these:
   \begin{eqnarray}
   \begin{picture}(160,-50)(0,20)
   \multiframe(50,25)(10.5,0){2}(10,10){a}{c}
   \multiframe(50,14.5)(10.5,0){2}(10,10){b}{d}
   \put(90,20){.}
   \end{picture}
   \label{BeBo:lambda}
   \end{eqnarray}
   First symmetrize over the first and third indices ($a$ and $c$), and over
   the second and fourth ($b$ and $d$):
   $$ \frac{1}{4}\,(T_{abcd}+T_{cbad}+T_{adcb}+T_{cdab})\,.$$
   Now antisymmetrize the result over the first and second indices ($a$ and $b$),
   and over the third and fourth ($c$ and $d$);\footnote{\label{BeBo:footnote}Here we adopt the
   convention
   that the second round of permutations interchanges indices with the same
   \emph{names}, rather than indices in the same \emph{positions} in the various terms.
   The opposite convention is tantamount to antisymmetrizing \emph{first}, which leads
   to a different, but mathematically isomorphic, development of the representation
   theory. The issue here is analogous to the distinction between space-fixed and
   body-fixed axes in the study of the rotation group (active or passive transformations).}
   dropping the combinatorial factor $\frac{1}{16}$, we get
   {\small{
   \begin{eqnarray}
    R_{ab|cd}&=&T_{abcd}+T_{cbad}+T_{adcb}+T_{cdab}-T_{bacd}-T_{cabd}-T_{bdca}-T_{cdba}
   \nonumber \\
    &-&T_{abdc}-T_{dbac}-T_{acdb}-T_{dcab}+T_{badc}+T_{dabc}+T_{bcda}+T_{dcba}\,.
   \nonumber
   \end{eqnarray}}}
  It is easy to check that $R$ possesses the symmetries of the Riemann tensor.
  There are two independent orders of its indices (e.g. $R_{ab|cd}$ and $R_{ac|bd}$),
  and applying any permutation
  to the indices produces some linear combination of those two basic objects.
  On the other hand, performing on $T$ the operations prescribed by the first
  tableau in (\ref{BeBo:4.3}) produces a different expression
  $P_{ac|bd}\,$, which, however,
  generates a two-dimensional representation of $\mathfrak{S}_4$ with the same abstract
  index structure as that generated by $R\,$. A non-standard tableau would also
  yield such a representation, but the tensors within it would be linear combinations
  of those already found.
  One finds
  {\small{
   \begin{eqnarray}
    P_{ac|bd}&=&T_{abcd}+T_{bacd}+T_{abdc}+T_{badc}-T_{cbad}-T_{bcad}-T_{cbda}-T_{bcda}
   \nonumber \\
             &-&T_{adcb}-T_{dacb}-T_{adbc}-T_{dabc}+T_{cdab}+T_{dcab}+T_{cdba}+T_{dcba}\,.
   \nonumber
   \end{eqnarray}}}
   As the reader may check, no linear combinations of $P$ can
   reproduce $R\,$.
   The objects $P_{ab|cd}$, $P_{ac|bd}$, $R_{ab|cd}$ and $R_{ac|bd}$
   are linearly independent.
   Although $R$ and $P$ are characterized by the same Young {\emph{diagram}},
   they are associated with different standard Young {\emph{tableaux}}
   and therefore span two {\emph{different}} (although equivalent) 
   irreducible representations of $\mathfrak{S}_n\,$. 
   Two representations may indeed be equivalent without being identical. 
   This happens in particular for the 
   irreducible decomposition of the regular representation of $\mathfrak{S}_n$ 
   where every irreducible representation appears with a multiplicity equal to its
   dimension. When the dimension of an $\mathfrak{S}_n$ irreducible 
   representation is $d>1\,$, then $d$ copies of that irreducible representation
   appear in the decomposition of the regular representation of $\mathfrak{S}_n$ 
   and all these $d$ representations are equivalent, although not identical.
  \vspace*{2mm}

  \noindent\textbf{Example:} Define a \emph{symmetrized Riemann tensor}
  (the \textsl{Jacobi tensor})
  by $J_{ad;bc}:=\frac{1}{2}\,(R_{ab|cd}+R_{ac|bd})\,$. It obeys $J_{ab;cd}=J_{ba;cd}=J_{ab;dc}\,$.
  Then it is easy to show that
  $R_{ab|cd}=\frac{2}{3}\,(J_{ad;bc}-J_{bd;ac})\,$. Thus the tensor $J$ has no fewer
  independent components and contains no less information than the tensor $R$,
  despite the extra symmetrization; $R$ is recovered from $J$ by an antisymmetrization.
  The tensors $R$ and $J$ are really the same tensor expressed with respect to different
  bases.
    \item[\bf{B}.] The \textsl{regular representation} of $\mathfrak{S}_n$ is the
    $n!$-dimensional
    representation obtained by letting $\mathfrak{S}_n$ act by left multiplication on the formal
    linear combinations of elements of $\mathfrak{S}_n\,$.
    [That is, one labels the basis vectors
    of $\mathbb{R}^{n!}$ by elements of $\mathfrak{S}_n$,
    defines that action of each permutation
    on the basis vectors in the natural way, and extends this definition to the whole
    space by linearity.] Equivalently, the vector space of the regular representation
    is the space of real-valued functions defined on $\mathfrak{S}_n\,$.
    [In general the regular representation is defined with complex scalars,
    but for $\mathfrak{S}_n$ it is sufficient to work with real coefficients.]

    \vspace{2mm}\noindent \textbf{Regular representation:}
    \textit{The regular representation contains
    every irreducible representation with a multiplicity equal to its dimension.
    Each Young diagram $\lambda$ corresponds to an irreducible
    representation of $\mathfrak{S}_n\,$.
    Its dimension and multiplicity are equal to the number
    of standard tableaux of diagram $\lambda\,$.}
\end{enumerate}

The symmetrization procedure described under \textbf{A.} can be
transcribed to the more abstract context \textbf{B.} to construct
a projection operator onto the subspace of $\mathbb{R}^{n!}$
supporting each representation. [The numerical coefficient needed
to normalize the tableau operation as a projection --- an operator
whose square is itself --- is not usually the same as that
accumulated from the individual symmetrization operations. For
example, to make $R_{abcd}$ into a projection of $T_{abcd}$, one
needs to divide by $12$, not $16$.] \vspace*{.2cm}

\vspace{2mm}\noindent\textbf{Example:} In (\ref{BeBo:4.2}), the
partition $\{4\}$ corresponds to the totally symmetric four-index
tensors, a one-dimensional space $V^{\mathfrak{S}_4}_{\{4\}}$.
Similarly, $\{1,1,1,1\}$ yields the totally antisymmetric tensors.
A generic rank-four tensor, $T_{abcd}$, can be decomposed into the
sum of its symmetric and antisymmetric parts, plus a remainder.
The theory we are expounding here tells how to decompose the
remainder further. The partition $\{2,2\}$ yields two independent
two-dimensional subrepresentations of the regular representation;
in more concrete terms, there are two independent pieces of
$T_{abcd}$ ($\frac{1}{12}\,R_{ab|cd}$ and
$\frac{1}{12}\,P_{ac|bd}$) constructed as described in connection
with (\ref{BeBo:4.3}). One of these ($R_{ab|cd}$) has exactly the
symmetries of the Riemann tensor; the other ($P_{ac|bd}$, coming
from the first tableau of (\ref{BeBo:4.3})) has the same abstract
symmetry as the Riemann tensor, but with the indices ordered
differently. Finally, each of the remaining partitions in
(\ref{BeBo:4.2}), i.e., $\{3,1\}$ and $\{2,1,1\}\,$, 
can be made into a standard tableau in three
different ways. Therefore, each of these two representations has
three separate pieces of $T$ corresponding to it, and each piece
is three-dimensional (has three independent index orders after its
symmetries are taken into account). Thus the total number of
independent tensors which can be formed from the irreducible parts
of $T_{abcd}$ by index permutations is $$
1^2+1^2+2^2+3^2+3^2=24=4! $$ which is simply the total number of
permutations of the indices of $T$ itself, as it must be.
\vspace*{3mm}

To state a formula for the dimension of an irreducible
representation $V^{\mathfrak{S}_n}_{\lambda}$ of
$\mathfrak{S}_n\,$, we need the concept of the hook length of a
given box in a Young diagram $\lambda\,$. The {\textsl{hook
length}} of a box in a Young diagram is the number of squares
directly below or directly to the right of the box, including the
box once:
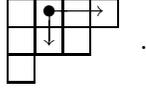
\begin{center}
\begin{picture}(80,30)(0,-15)
\multiframe(0,0)(10.5,0){1}(10,10){}
\multiframe(10.5,0)(10.5,0){1}(10,10){}
\multiframe(21,0)(10.5,0){1}(10,10){}
\multiframe(31.5,0)(10.5,0){1}(10,10){}
\multiframe(0,-10.5)(10.5,0){1}(10,10){}
\multiframe(10.5,-10.5)(10.5,0){1}(10,10){}
\multiframe(21,-10.5)(10.5,0){1}(10,10){}
\multiframe(0,-21)(10.5,0){1}(10,10){}
\put(13,-5){$\downarrow$}\put(19,3){$\longrightarrow$}\put(15,3){$-$}
\put(14.3,-3){$|$} \put(50,-8.5){.} \put(13,3){$\bullet$}
\end{picture}
\end{center}
\vspace{2mm}\noindent\textbf{Example:} In the following diagram, each box is labeled by its hook length:
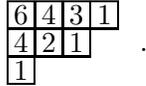
\begin{center}
\begin{picture}(80,30)(0,-20)
\multiframe(0,0)(10.5,0){1}(10,10){6}
\multiframe(10.5,0)(10.5,0){1}(10,10){4}
\multiframe(21,0)(10.5,0){1}(10,10){3}
\multiframe(31.5,0)(10.5,0){1}(10,10){1}
\multiframe(0,-10.5)(10.5,0){1}(10,10){4}
\multiframe(10.5,-10.5)(10.5,0){1}(10,10){2}
\multiframe(21,-10.5)(10.5,0){1}(10,10){1}
\multiframe(0,-21)(10.5,0){1}(10,10){1} \put(50,-8.5){.}
\end{picture}
\end{center}

\vspace{2mm}One then has the following {\textsl{hook length formula}} for the
dimension of the representation $V^{\mathfrak{S}_n}_{\lambda}$ of
$\mathfrak{S}_n$ corresponding to the Young diagram $\lambda\,$:
\begin{eqnarray}
\dim V^{\mathfrak{S}_n}_{\lambda}=\frac{n!}{\prod (\mbox{ hook
lengths} )}\,. \label{BeBo:4.12}
\end{eqnarray}

\vspace{2mm}\noindent\textbf{Remark:} Note carefully that the
``dimension'' we have been discussing up to now is the number of
independent \emph{index orders} of a tensor, not the number of
independent \emph{components} when the tensor is realized
geometrically with respect to a particular underlying vector space
or manifold. The latter number depends on the dimension (say $D$)
of that underlying space, while the former is independent of $D$
(so long as $D$ is sufficiently large, as we tacitly assume in
generic discussions). For example, the number of components of an
antisymmetric two-index tensor is $\frac{D(D-1)}{2}\,$, but the
number of its index orders is always $1$, except in dimension
$D=1$ where no non-zero antisymmetric tensors exist at all.
\vspace*{.3cm}

\subsection{General linear group}
\label{BeBo:genlingr}
%

We now turn to the representation theory of the general linear and
orthogonal groups, where the (spacetime) dimension $D$ plays a key
role. The theory of partitions and of the representations of the
permutation groups is the foundation on which this topic is built.

Let $\{v_a\}$ represent a generic element of $\mathbb{R}^{D*}$ (or of
the cotangent space at a point of a $D$-dimensional manifold). The
action of non-singular linear operators on this space gives a
$D$-dimensional irreducible representation $V\cong \mathbb{R}^{D*}$ 
of the {general linear group} $GL(D)\,$; indeed, this representation defines the
group itself. The rank-two tensors, $\{T_{ab}\}$, carry a larger
representation of $GL(D)$ ($V\otimes V$, of dimension $D^2$),
where the group elements act on the two indices simultaneously.
The latter representation is reducible: it decomposes into the
subspace of symmetric and antisymmetric rank-two tensors $V\otimes
V\cong (V\odot V) \oplus (V\wedge V)$, of respective dimensions
$\frac{D(D+1)}{2}\,$ and $\frac{D(D-1)}{2}\,$. Similarly, the
tensor representation of rank $n$, $V^{\otimes n}$, decomposes
into irreducible representations of $GL(D)$ which are associated
with the irreducible representations of $\mathfrak{S}_n$ acting on
the indices, which in turn are labeled by the partitions of $n\,$,
hence by Young diagrams. Young diagrams with more than $D$ rows do
not contribute [if $\lambda$ is a partition of $n$ into more than
$D$ parts, then the associated index symmetrization of a
$D$-dimensional rank-$n$ tensor yields an expression that vanishes
identically; in particular, there are no non-zero totally
antisymmetric rank-$n$ tensors if $n>D\,$].

More precisely, let $\lambda$ be a Young \emph{tableau}. The
\textsl{Schur module} $V_\lambda^{GL(D)}$ is the vector space of
all rank-$n$ tensors $\tilde T$ in $V^{\otimes n}$ such that:
\begin{quote}

$(i)$ the tensor $\tilde T$ is completely antisymmetric in the
entries of each column of $\lambda\,$,

$(ii)$ complete antisymmetrization of $\tilde T$ in the entries of
a column of $\lambda$ and another entry of $\lambda$ that is on
the right-hand side of the column vanishes.
\end{quote}
This construction is equivalent to the construction {\bf{A.}}

\vspace{2mm}\noindent\textbf{Example:} Associated with the Young
tableau (\ref{BeBo:lambda}), the tensor $R_{ab|cd}$ introduced in
the subsection \ref{BeBo:permutationgr} obeys to the conditions
(i) and (ii): $R_{ab|cd}=-R_{ba|cd}=-R_{ab|dc}$ and
$R_{ab|cd}+R_{bc|ad}+R_{ca|bd}=0\,$.\vspace{2mm}

As explained in the footnote \ref{BeBo:footnote}, if one
interchanges everywhere in the previous constructions the words
``symmetric" and ``antisymmetric," then the (reducible)
representation spaces characterized by the same Young
{\emph{diagram}} [but not by the same Young {\emph{tableau}}] are
isomorphic and the conditions (i)-(ii) must be replaced with:
\begin{quote}

$(a)$ the tensor is completely (or totally) symmetric in the entries of each
column of $\lambda$,

$(b)$ complete symmetrization of the tensor in the entries of a
row of $\lambda$ and another entry of $\lambda$ that sits in a
lower row vanishes.
\end{quote}

\vspace{2mm}\noindent\textbf{Example:} Taking the standard Young
tableau (\ref{BeBo:lambda}) and constructing, following the
``manifestly symmetric convention'', the irreducible tensor
associated with it, one obtains a tensor ${\cal{R}}$ with the same
abstract index symmetries as $J$ [{\it i.e.} obeying the
constraints $(a)$ and $(b)$] but which is however linearly
independent from $J\,$, thence linearly independent from $R$
alone. The tensor ${\cal{R}}$ can be expressed as a linear
combination of {\emph{both}} $R$ {\emph{and}} $P\,$. Similarly,
taking the first standard Young tableau in (\ref{BeBo:4.3}) and
following the manifestly symmetric convention, one obtains a
tensor $\cal P$ obeying $(a)$ and $(b)$. This tensor is linearly
independent from $P$ alone as it is a linear combination of
\emph{both} $P$ and $R\,$. Summarizing, associated with the Young
{\emph{diagram}} $\{2,2\}$ we have the (reducible) representation
space spanned by either $\{R,P\}$ in the manifestly antisymmetric
convention or by $\{\cal{R},\cal{P}\}$ in the manifestly symmetric
convention.

\vspace{2mm}\noindent\textbf{Remarks:}

\noindent $\bullet$ An important point to note is that, by the
previous construction featuring irreducible tensors with definite
symmetry properties, one generates essentially \emph{all}
the finite-dimensional irreducible representations of 
$GL(D,\mathbb{R})\,$. 
To be more precise, 
$GL(D,\mathbb{R})\,$ tensors can be of type $(p,q)\,$, 
i.e., having $p$ contravariant indices and $q$ covariant ones. 
The exhaustive list of finite-dimensional irreducible representations 
of $GL(D,\mathbb{R})\,$ is provided by $(p,q)$-type tensors 
characterised by a pair of Young tableaux of rank $p$ and $q\,$, 
respectively, and such that the contraction of any covariant index
with a contravariant one gives zero identically.  
See e.g. Chapter 13 of \cite{Tung} for more details.

\noindent $\bullet$ In order to make contact with an alternative
road to the representation theory of $GL(D)$, one says that the
irreducible representation
$\Gamma_{\lambda^1\,\ldots\,\lambda^{D-1}}$ of
${\mathfrak{sl}}(D,\mathbb C)\equiv A_{D-1}$ with highest weight
$\Lambda=\lambda^1\Lambda_{(1)}+\lambda^2\Lambda_{(2)}+\ldots+
\lambda^{D-1}\Lambda_{(D-1)}$ [see \textit{e.g.} the Part II of
the lecture notes \cite{BeBo:Liealgs} for definitions and
notations] is obtained by applying the Schur functor ${\mathbb
S}_{\lambda}$ [\textit{i.e.} the construction presented above] to
the standard representation $V$, where the Young diagram is
\begin{eqnarray}
\lambda=\{\lambda^1+\ldots+\lambda^{D-1}\,,\,\lambda^2+\ldots+\lambda^{D-1}\,,\,
\ldots\,,\lambda^{D-1}\,,\,0\}\,. \nonumber
\end{eqnarray}
In terms of the Young diagram for $\lambda\,$, the Dynkin labels
$\lambda^a$ ($1\leqslant a \leqslant D-1$) are the differences of
lengths of rows: $\lambda^a=\lambda_a-\lambda_{a+1}\,$.

\vspace{2mm}\noindent\textbf{Example:} If $D=6$, then
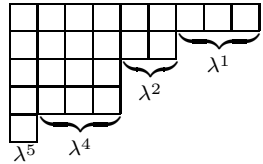
\begin{center}
\begin{picture}(0,60)(30,-35)
\multiframe(0,10)(10.5,0){1}(10,10){}
\multiframe(10.5,10)(10.5,0){1}(10,10){}
\multiframe(21,10)(10.5,0){1}(10,10){}
\multiframe(31.5,10)(10.5,0){1}(10,10){}
\multiframe(42,10)(10.5,0){1}(10,10){}
\multiframe(52.5,10)(10.5,0){1}(10,10){}
\multiframe(63,10)(10.5,0){1}(10,10){}
\multiframe(73.5,10)(10.5,0){1}(10,10){}
\multiframe(84,10)(10.5,0){1}(10,10){}
\multiframe(0,-0.5)(10.5,0){1}(10,10){}
\multiframe(10.5,-0.5)(10.5,0){1}(10,10){}
\multiframe(21,-0.5)(10.5,0){1}(10,10){}
\multiframe(31.5,-0.5)(10.5,0){1}(10,10){}
\multiframe(42,-0.5)(10.5,0){1}(10,10){}
\multiframe(52.5,-0.5)(10.5,0){1}(10,10){}
\multiframe(0,-11)(10.5,0){1}(10,10){}
\multiframe(10.5,-11)(10.5,0){1}(10,10){}
\multiframe(21,-11)(10.5,0){1}(10,10){}
\multiframe(31.5,-11)(10.5,0){1}(10,10){}
\multiframe(0,-21.5)(10.5,0){1}(10,10){}
\multiframe(10.5,-21.5)(10.5,0){1}(10,10){}
\multiframe(21,-21.5)(10.5,0){1}(10,10){}
\multiframe(31.5,-21.5)(10.5,0){1}(10,10){}
\multiframe(0,-32)(10.5,0){1}(10,10){}
\put(64,9.5){$\underbrace{\hspace*{30pt}}_{\lambda^1}$}
\put(43,-1){$\underbrace{\hspace*{20pt}}_{\lambda^2}$}
\put(11.5,-22){$\underbrace{\hspace*{30pt}}_{\lambda^4}$}
\put(1,-37){${}_{\lambda^5}$}
\end{picture}
\end{center}
is the Young diagram corresponding to the irrep
$\Gamma_{3,2,0,3,1}$ of $A_5\equiv
{\mathfrak{sl}}(6,\mathbb{C})\,$. \vspace*{.4cm}

The dimension of the representation $V^{GL(D)}_{\lambda}$ of
$GL(D)$ corresponding to the Young diagram $\lambda\,$ is:
\begin{eqnarray}
\dim V^{GL(D)}_{\lambda}=\,\prod\,\frac{D-\mbox{row}+\mbox{column}
}{ \mbox{ hook length} }\,, \label{BeBo:dimension}
\end{eqnarray}
where the product is over the $n$ boxes while ``row" and ``column"
respectively give the place of the corresponding box. As was
underlined before, the formula (\ref{BeBo:dimension}) is distinct
from the hook length formula (\ref{BeBo:4.12}).

\vspace{2mm}\noindent\textbf{Examples:}

\noindent $\bullet$ In the following diagram
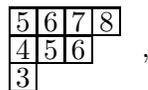
\begin{center}
\begin{picture}(80,30)(0,-20)
\multiframe(0,0)(10.5,0){1}(10,10){5}
\multiframe(10.5,0)(10.5,0){1}(10,10){6}
\multiframe(21,0)(10.5,0){1}(10,10){7}
\multiframe(31.5,0)(10.5,0){1}(10,10){8}
\multiframe(0,-10.5)(10.5,0){1}(10,10){4}
\multiframe(10.5,-10.5)(10.5,0){1}(10,10){5}
\multiframe(21,-10.5)(10.5,0){1}(10,10){6}
\multiframe(0,-21)(10.5,0){1}(10,10){3} \put(50,-8.5){,}
\end{picture}
\end{center}
each box is labeled by its value in the numerator of
(\ref{BeBo:dimension}) for $D=5$. Observe that, for the
corresponding diagram $\lambda\,$, $\dim
V^{GL(5)}_{\lambda}=1050\neq 70=\dim
V^{\mathfrak{S}_8}_{\lambda}\,$.

\noindent $\bullet$ The space of (anti)symmetric tensors of $V$ of
rank $n$ are denoted by $\odot^n(V)$ (respectively,
$\wedge^n(V)$). It carries an irreducible representation of
$GL(D)$ labeled by a Young diagram made of one row (respectively,
column) of length $n\,$. The dimensions
\begin{equation}
\dim\odot^n(V)= \Big(\begin{array}{c}
  D+n-1 \\
  n
\end{array}\Big)\,,
\quad\quad\dim\wedge^n(V)=\Big(\begin{array}{c}
  D \\
  n
\end{array}\Big)\,,
\label{BeBo:(anti)sym}
\end{equation}
are easily computed from the formula (\ref{BeBo:dimension}) and
reproduce the standard results obtained from combinatorial
arguments.\vspace{2mm}

If $T_1$ and $T_2$ are tensors of ranks $n_1$ and $n_2\,$, 
respectively, then
their tensor product is a tensor of rank $n_1+n_2\,$. Each factor $T_j$
transforms under index permutation according to some
representation of $\mathfrak{S}_{n_j}\,$, and under linear
transformation by the corresponding representation of $GL(D)\,$.
It follows immediately that the tensor product  $T_1\otimes T_2$
transforms as some representation of $\mathfrak{S}_{n_1}\times
\mathfrak{S}_{n_2}\,$. This induces a representation of the
full permutation group $\mathfrak{S}_{n_1+n_2}$ which is
associated with a corresponding representation of $GL(D)\,$. 
It is possible to reduce these
last two representations into a sum of irreducible ones. We may
assume that the factor representations are irreducible, since the
original tensors $T_j$ could have been broken into irreducible
parts at the outset. \vspace*{.2cm}

\vspace{2mm}\noindent\textbf{Littlewood--Richardson rule:} The
decomposition of an ``outer product'' $\mu\cdot\nu$ of irreducible
representations $\mu$ and $\nu$ of $\mathfrak{S}_{n_1}$ and
$\mathfrak{S}_{n_2}$, respectively, into irreducible
representations of $\mathfrak{S}_{n_1+n_2}$ can be determined by
means of the following algorithm involving Young diagrams. The
product is commutative, so it does not matter which factor is
regarded as the ``right-hand'' one. [In practice, on should choose
the simpler Young diagram for that role.]
\begin{itemize}
    \item[(I)] Label each box in the top row of the right-hand diagram, $\nu$,
    by ``$a$'', each box in the second row by ``$b$'', etc.
    \item[(II)] Add the labeled boxes of $\nu$ to the left-hand diagram $\mu$,
    one at a time, first the $a$s, then the $b$s, ..., subject to these constraints:
  \begin{enumerate}
      \item[(A)] No two boxes in the same column are labeled with the same letter;
      \item[(B)] At all stages the result is a legitimate Young diagram;
      \item[(C)] At each stage, if the letters are read right-to-left along the rows,
      from top to bottom, one never encounters more $b$s than $a$s, more $c$s than $b$s, etc.
  \end{enumerate}
  \item[(III)] Each of the distinct diagrams constructed in this way specifies an
  irreducible subrepresentation $\lambda$, appearing in the decomposition of
  the outer product. The same labeled Young diagram may arise in more than one way;
  the multiplicity of that representation must be counted accordingly.
\end{itemize}

\vspace{2mm}\noindent\textbf{Remarks:}

\noindent $\bullet$ This rule enables products of
\emph{distinct} tensors to be decomposed. When the factors are the
same tensor, the list is further restricted by the requirement of
symmetry under interchange of the factors. This is the problem of
\textsl{plethysm}, whose solution requires more complicated
techniques than the Littlewood--Richardson rule.

\noindent $\bullet$ Representations with too
many parts (columns of length greater than $D$) must be deleted
from the list of subrepresentations of the $GL(D)$. [If
irreducible representations of the special linear group $SL(D)$
are considered instead, every column of length $D$ must be removed
from the corresponding Young diagram.]
  
\subsection{Orthogonal group}
\label{BeBo:ortho}
%

It remains to consider index contractions. Up to now we
considered only covariant tensors, because in the intended
application there is a metric tensor which serves to relate
contravariant and covariant tensors. Contractions are mediated by
this metric. Implicitly, therefore, one is restricting the
symmetry group of the problem from the general linear group to the
subgroup that leaves the metric tensor invariant, the {orthogonal
group} $O(D)\,$. [If the metric has indefinite signature, the true
symmetry group is a non-compact analogue of the orthogonal group,
such as the Lorentz group. This does not affect the relevant
aspects of the \textit{finite-dimensional} representation theory.]
Each irreducible $GL(D)$ representation $V^{GL(D)}_\lambda$
decomposes into irreducible $O(D)$ representations $V_\nu^{O(D)}$,
labeled by Young diagrams $\nu$ obtained by removing an even
number of boxes from $\lambda\,$. The \textsl{branching rule} for
this process involves a sort of inverse of the
Littlewood--Richardson rule:

\vspace{2mm}\noindent\textbf{Restriction from $GL(D)$ to $O(D)$:}
\textit{The irreps of $GL(D)$ may be reduced to direct sums of
irreps of $O(D)$ by extracting all possible trace terms formed by
contraction with products of the metric tensor and its
inverse.}\vspace{1mm}

The reduction is given by the branching rule for 
$GL(D)\downarrow O(D)$:
\begin{equation}
V_\lambda^{GL(D)}  =  V_{\lambda/\Delta}^{O(D)}\equiv
V_\lambda^{O(D)}\oplus V_{\lambda/\{2\}}^{O(D)}\oplus V_{
\lambda/\{4\}}^{O(D)}\oplus
V_{\lambda/\{2,2\}}^{O(D)}\oplus\,\ldots \label{BeBo:reduction}
\end{equation}
where $\Delta$ is the formal infinite sum \cite{BeBo:King}
$$\Delta=\,1\,+\begin{picture}(30,15)(-5,2)
\multiframe(0,0)(10.5,0){2}(10,10){}{}
\end{picture}
+
\begin{picture}(55,15)(-5,2)
\multiframe(0,0)(10.5,0){4}(10,10){}{}{}{}
\end{picture}
+
\begin{picture}(30,15)(-5,2)
\multiframe(0,5)(10.5,0){2}(10,10){}{}
\multiframe(0,-5)(10.5,0){2}(10,10){}{}
\end{picture}
+\ldots
$$ corresponding to the sum of all possible plethysms of the metric tensor,
and where $\lambda/\mu$ means the sum of the Young diagrams $\nu$
such that $\nu\cdot \mu$ contains $\lambda$ according to the
Littlewood--Richardson rule (with the corresponding multiplicity).

\vspace{2mm}\noindent\textbf{Examples:}

\noindent$\bullet$ The $GL(D)$ irreducible representation labeled
by the Young diagram $\{2,2\}$ decomposes with respect to $O(D)$
according to the direct sum
$\{2,2\}/\Delta=\{2,2\}+\{2,0\}+\{0,0\}$ which corresponds to the
decomposition of the Riemann tensor into the Weyl tensor, the
traceless part of the Ricci tensor and the scalar curvature,
respectively.

\noindent$\bullet$ The $GL(D)$ irreducible representation labeled
by the Young diagram $\{n\}$ decomposes with respect to $O(D)$
according to the direct sum
$\{n\}/\Delta=\{n\}+\{n-2\}+\{n-4\}+\,\ldots\,$, corresponding to
the decomposition of a completely symmetric tensor or rank $n$
into its traceless part, the traceless part of its trace,
\textit{etc}. This provides an alternative proof of the obvious
fact that the number of independent components of a traceless
symmetric tensor of rank $n$ is equal to the number of independent
components of a symmetric tensor of rank $n$ minus the number of
independent components of a symmetric tensor of rank $n-2$ (its
trace): $\dim V_{\{n\}}^{O(D-2)}=\dim V_{\{n\}}^{GL(D)}-\dim
V_{\{n-2\}}^{GL(D)}\,$. Using the formula (\ref{BeBo:(anti)sym})
allows to show that
\begin{equation}
\dim V_{\{n\}}^{O(D)}=\frac{(D+2n-2)(D+n-3)!}{n!(D-2)!}\,.
\label{BeBo:dimsym}
\end{equation}
The very useful formula (\ref{BeBo:dimsym}) contains as a
particular case the well-known fact that all the traceless
symmetric tensorial representations of $O(2)$ are two-dimensional
(indeed, any UIR of an Abelian group is of complex dimension one).
Moreover, the traceless symmetric tensorial representations of
rank $n$ of the rotation group $O(3)$ are the well-known integer
spin representations of dimension equal to $2n+1\,$.\vspace*{.2cm}

The following theorem is important (see e.g. the first
reference of~\cite{BeBo:Hamermesh}):

\newpage\noindent\textbf{Vanishing irreps for
(pseudo-)orthogonal groups:} \textit{Whenever the sum of the
lengths of the first two columns of a Young diagram $\lambda$ is
greater than $D=p+q\,$, then the irreducible representation of 
$O(p,q)$ labeled by $\lambda$ is identically zero.}

\vspace{2mm}Young diagrams such that the sum of the lengths of the
first two columns does not exceed $D$ are said to be
{\textsl{allowed}}. \vspace*{.4cm}

\noindent {\textbf{Finite-dimensional irreps of
(pseudo-)orthogonal groups:}} \textit{Each non-zero
finite-dimensional irreducible representation of $O(p,q)$
is isomorphic to a completely traceless tensorial representation,
the symmetry properties of which are labeled by an allowed Young
diagram $\lambda\,$.}\vspace{2mm}

The dimension of the tensorial irrep is determined by the
following rule due to King \cite{BeBo:KING}:
\begin{itemize}
  \item[($\alpha$)] The numbers $D-1$, $D-3$, $D-5$, $\ldots\,$,
  $D-2r+1$ are placed in the end boxes of the $1$st, $2$nd, $3$rd,
  $\ldots$, $r$th rows of the diagram $\lambda\,$. A labeled Young
  diagram of $n$ numbers is then constructed by inserting in the
  remaining boxes of the diagram, numbers which increase by one in
  passing from one box to its left-hand neighbor.
  \item[($\beta$)] This labeled Young diagram is extended to the
  limit of the triangular Young diagram $\tau$ of $r$ rows. This produces
  a Young diagram $\widetilde{\lambda}$ the $a$th row of which has length equal
  the maximum between the two integers $\tau_a=r-a+1$ and $\lambda_a$.
  \item[($\gamma$)] The series of numbers in any row of the
  Young diagram $\widetilde{\lambda}$
  is then extended by inserting in the remaining boxes of the diagram,
  numbers which decrease by one in
  passing from one box to its right-hand neighbor. The resulting
  numbers will be called the ``King length."
  \item[($\delta$)] The row lengths $\lambda_1$, $\lambda_2$,
  $\ldots$, $\lambda_r$ are then added to all of the numbers of
  the Young diagram $\widetilde{\lambda}$ which lie on
  lines of unit slope passing through the first box of the $1$st,
  $2$nd, $\ldots$, $r$th rows, respectively, of the Young diagram
  $\lambda\,$.
\end{itemize}
The dimension is equal to the product of the integers in the
resulting labeled Young diagram $\widetilde{\lambda}$ divided by
the product of
\begin{itemize}
  \item[-] the hook length of each box of $\lambda$, and of
  \item[-] the King length of each box of $\widetilde{\lambda}$
  outside $\lambda\,$.
\end{itemize}

\vspace{2mm}\noindent\textbf{Examples:}\vspace{1mm}

\noindent $\bullet$ In the following diagram, allowed for $D=5$,
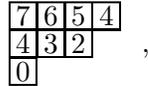
\begin{center}
\begin{picture}(80,30)(0,-20)
\multiframe(0,0)(10.5,0){1}(10,10){7}
\multiframe(10.5,0)(10.5,0){1}(10,10){6}
\multiframe(21,0)(10.5,0){1}(10,10){5}
\multiframe(31.5,0)(10.5,0){1}(10,10){4}
\multiframe(0,-10.5)(10.5,0){1}(10,10){4}
\multiframe(10.5,-10.5)(10.5,0){1}(10,10){3}
\multiframe(21,-10.5)(10.5,0){1}(10,10){2}
\multiframe(0,-21)(10.5,0){1}(10,10){0} \put(50,-8.5){,}
\end{picture}
\end{center}
each box is labeled by its King length, while in the diagram
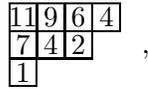
\begin{center}
\begin{picture}(80,30)(0,-20)
\multiframe(0,0)(10.5,0){1}(10,10){11}
\multiframe(10.5,0)(10.5,0){1}(10,10){9}
\multiframe(21,0)(10.5,0){1}(10,10){6}
\multiframe(31.5,0)(10.5,0){1}(10,10){4}
\multiframe(0,-10.5)(10.5,0){1}(10,10){7}
\multiframe(10.5,-10.5)(10.5,0){1}(10,10){4}
\multiframe(21,-10.5)(10.5,0){1}(10,10){2}
\multiframe(0,-21)(10.5,0){1}(10,10){1} \put(50,-8.5){,}
\end{picture}
\end{center}
each box is labeled by the number obtained at the very end of
King's rule. Observe that, for the corresponding diagram
$\lambda\,$, it was not necessary to perform the steps
($\beta$)-($\gamma$) and that, $\dim
V^{O(5)}_{\lambda}=231<1050=\dim V^{GL(5)}_{\lambda}\,$.\vspace{1mm}

\noindent $\bullet$ In the following Young diagram $\lambda=\{2,2,1\}\,$, allowed for $D=5$,
\begin{center}
\begin{picture}(80,30)(0,-20)
\multiframe(0,0)(10.5,0){1}(10,10){5}
\multiframe(10.5,0)(10.5,0){1}(10,10){4}
\multiframe(0,-10.5)(10.5,0){1}(10,10){3}
\multiframe(10.5,-10.5)(10.5,0){1}(10,10){2}
\multiframe(0,-21)(10.5,0){1}(10,10){0} \put(50,-8.5){,}
\end{picture}
\end{center}
each box is labeled by the number obtained after step $(\alpha)\,$. 
The step $(\beta)$ is now necessary and gives the Young diagram $\widetilde{\lambda}=\{3,2,1\}\,$. 
At the end of steps $(\gamma)$ and $(\delta)$, respectively, the result is
\begin{center}
\begin{picture}(80,30)(0,-20)
\put(-30,-8.5){$\stackrel{(\gamma)}{\longrightarrow}$}
\multiframe(0,0)(10.5,0){1}(10,10){5}
\multiframe(10.5,0)(10.5,0){1}(10,10){4}
\multiframe(21,0)(10.5,0){1}(10,10){3}
\multiframe(0,-10.5)(10.5,0){1}(10,10){3}
\multiframe(10.5,-10.5)(10.5,0){1}(10,10){2}
\multiframe(0,-21)(10.5,0){1}(10,10){0} 
\put(40,-8.5){$\stackrel{(\delta)}{\longrightarrow}$}
\multiframe(70,0)(10.5,0){1}(10,10){7}
\multiframe(80.5,0)(10.5,0){1}(10,10){6}
\multiframe(91,0)(10.5,0){1}(10,10){4}
\multiframe(70,-10.5)(10.5,0){1}(10,10){5}
\multiframe(80.5,-10.5)(10.5,0){1}(10,10){3}
\multiframe(70,-21)(10.5,0){1}(10,10){1} \put(120,-8.5){,}
\end{picture}
\end{center}
so that $\dim V^{O(5)}_{\lambda}=$
$\frac{7\cdot 6\cdot 5\cdot 4\cdot 3}{(4\cdot 3\cdot 2)\cdot(3)}\,=35<75$
$=\dim V^{GL(5)}_{\lambda}\,$.

\noindent $\bullet$ The space of traceless symmetric tensors of
$V$ of rank $n$ carries an irreducible representation of $O(D)$
labeled by a Young diagram made of one row of length $n\,$ for
which the dimension (\ref{BeBo:dimsym}) is easily reproduced from
the King rule, since the rules ($\beta$)-($\gamma$) may be omitted

\noindent $\bullet$ Computing the number of components of the Weyl 
tensor and of a symmetric, traceless, rank-two tensor in $D=4$ 
dimensions, enables one to give the decomposition 
$\{2,2\}/\Delta=\{2,2\}+\{2,0\}+\{0,0\}$ of the Riemann tensor 
into the Weyl tensor, the traceless part of the Ricci tensor and the 
scalar curvature, respectively, in terms of the corresponding dimensions. 
This gives the well-known result $20= 10 + 9 + 1\,$. 
\vspace{2mm}

\noindent {\textbf{Unitary irreps of orthogonal groups:}}
\textit{Each non-zero inequivalent UIR of $O(D)$ corresponds to an
allowed Young diagram $\lambda\,$, and conversely.}\vspace{2mm}

\vspace{2mm}\noindent\underline{Proof:} The orthogonal group is
compact, thence any UIR is finite-dimensional (see Subsection
\ref{BeBo:ABC}). Furthermore, any finite-dimensional irrep of the
orthogonal group is labeled by an allowed Young diagram. Moreover,
an important result is that any finite-dimensional representation
may be endowed with a sesquilinear form which makes it
unitary.\qed

\vspace{3mm}The quadratic Casimir operator of the orthogonal
algebra $\mathfrak{so}(D)$ presented by its generators and its
commutation relations
\begin{equation}i\,[M_{\mu\nu},M_{\rho\sigma}] =
\delta_{\nu\rho}M_{\mu\sigma}-\delta_{\mu\rho}M_{\nu\sigma}-\delta_{\sigma\mu}M_{\rho\nu}
+\delta_{\sigma\nu}M_{\rho\mu}\,\label{BeBo:sodcomrel}
\end{equation}
is the sum of square of the generators (similarly to the
definition (\ref{BeBo:QuadratiC}) for $\mathfrak{so}(D-1,1)$ since
these two \textit{complex} algebras are isomorphic). Its
eigenvalue on a finite-dimensional irrep labeled by an allowed
Young diagram $\lambda= \{\lambda_1,\lambda_2,\ldots,\lambda_r\}$
is given in the subsection 9.4.C of
\cite{BeBo:Barut}:
\begin{equation}
\Big[\,\,{\cal
C}_2\Big(\mathfrak{so}(D)\Big)\,-\,\sum_{a=1}^r\lambda_a(\lambda_a+D-2a)\,\Big]\,V_\lambda^{O(D)}=0\,.
\label{BeBo:quadrat}
\end{equation}

\vspace{2mm}\noindent\textbf{Examples:}

\noindent$\bullet$ The UIRs of the Abelian group $O(2)\cong U(1)$ are
labeled by one integer only, which is the eigenvalue of the single
generator on the irrep, say $h\in\mathbb Z\,$. The only allowed
Young diagrams are made of a single row of length equal to the
non-negative integer $s=|h|\,$. The traceless symmetric tensorial
representations of $O(2)$ are two-dimensional, the sum of the two
irreps labeled by $h=\pm s\,$. The formula (\ref{BeBo:quadrat})
with $D=2\,$, $r=1$ and $\lambda_1=s$ gives the obvious eigenvalue
$s^2\,$, since the quadratic Casimir operator of the rotation
group $O(2)$ is equal to the square of the single generator.

\noindent$\bullet$ The quadratic Casimir operator of the rotation
group $O(3)$ is the square of the angular momentum. The irrep of
$O(3)$ with spin $s\in\mathbb N$ is labeled by the allowed Young
diagram made of a single row of length equal to the integer $s\,$.
The formula (\ref{BeBo:quadrat}) with $D=3\,$, $r=1$ and
$\lambda_1=s$ gives the celebrated eigenvalue $s(s+1)\,$.

\noindent$\bullet$ The irrep of $O(D)$ carried by the space of
traceless symmetric tensors of rank $n$ is labeled by the allowed
Young diagram $\{n\}$ made of a single row of length equal to an
integer $n\,$. The formula (\ref{BeBo:quadrat}) with $r=1$ and
$\lambda_1=n$ gives the eigenvalue $n(n+D-2)$ for the quadratic
Casimir operator.

\vspace{2mm}The following branching rule is extremely useful in
the process of dimensional reduction.

\vspace{2mm}\noindent\textbf{Restriction from $GL(D)$ to
$GL(D-1)$:} \textit{The restriction to the subgroup
$GL(D-1)\subset GL(D)$ of a finite-dimensional irrep of $GL(D)$
determined by the Young diagram $\lambda$ contains each irrep of
$GL(D-1)$ labeled by Young diagrams $\mu$ such that
$$\lambda_1\geqslant\mu_1\geqslant\lambda_2\geqslant\mu_2
\geqslant\ldots\geqslant\mu_{r-1}\geqslant\lambda_r\geqslant\mu_r\geqslant
0\,,
$$ with multiplicity one. The same theorem holds for the
restriction $O(D)\downarrow O(D-1)$ where $\lambda$ is an allowed
Young diagram.}\vspace{1mm}

These rules are discussed in the section 8.8.A of
\cite{BeBo:Barut}. They may be summarised in the following
branching rule for $GL(D)\downarrow GL(D-1)$,
\begin{equation}
V_\lambda^{GL(D)}  =  V_{\lambda/\Sigma}^{GL(D-1)}\equiv
V_\lambda^{GL(D-1)}\oplus V_{\lambda/\{1\}}^{GL(D-1)}\oplus V_{
\lambda/\{2\}}^{GL(D-1)}\oplus
V_{\lambda/\{3\}}^{GL(D-1)}\oplus\,\ldots \label{BeBo:reduction2}
\end{equation}
where $\Sigma$ is the formal infinite sum of all Young diagrams
made of a single row.

\vspace{2mm}\noindent\textbf{Example:} The branching rule applied
to symmetric irrep labeled by a Young diagram $\{n\}$ made of one
row of length $n$ gives as a result:
$$\{n\}/\Sigma\,=\,\{\,n\,\,\}+\{n-1\}+\{n-2\}+\ldots+\{1\}+\{0\}\,.$$
This implies the obvious fact that a completely symmetric tensor
of rank $n$ whose indices run over $D$ values may be decomposed as 
a sum of completely symmetric tensors of rank $n$, $n-1$, $\ldots\,$, 
$1$, $0$ whose indices run over $D-1$ values. 
A non-trivial instance of the branching rule for $O(D)\downarrow O(D-1)$
is that the same result is true for \textit{traceless} symmetric tensors
as well.

\subsection{Auxiliary variables}
\label{BeBo:auxiliary}
%

Let $\lambda$ be a Young diagram with $s$ columns and $r$ rows.

The Schur module $V_\lambda^{GL(D)}$ in the ``manifestly
antisymmetric convention" can be built \textit{via} a convenient 
construction in terms of polynomials in $s\times D$ graded variables 
satisfying appropriate conditions. More precisely, the vector space
$V_\lambda^{GL(D)}$ is isomorphic to a subspace of the associative algebra
\begin{eqnarray}
{\cal A}
= (\otimes^{s}\wedge \mathbb{R}^{D*})\otimes C^{\infty}(\mathbb{R}^{D})
= \otimes^{s}_{C^{\infty}(\mathbb{R}^{D})}\Omega(\mathbb{R}^{D}) 
\end{eqnarray}
of $s$ tensor products of antisymmetric forms. 
The elements of ${\cal A}
$ are called \textsl{multiforms} \cite{DuboisViolette:2001jk}.

The $D$ generators of the
$I$th factor (where $I \in \{1,2,...,s\}$)  
$\mathbb{R}^{D*}$ in $(\otimes^{s}\wedge \mathbb{R}^{D*})$ 
are written $d_{_I}x^\mu\,$, where $\,\mu\in\{0,1,\ldots,D-1\}\,$. 
By definition, the multiform algebra $\cal A$ is
presented by the graded commutation relations
\begin{eqnarray}
d_{_I}x^\mu \,d_{_J}x^\nu\,=\,(-)^{\delta_{_{IJ}}}\,d_{_J}x^\nu\,
d_{_I}x^\mu\,, \label{BeBo:presented}
\end{eqnarray}
where the wedge  products are not written explicitly.
The condition (i) of Subsection \ref{BeBo:genlingr} is
automatically verified for any element
$\Phi\in{\cal A}
$ due to the fact that the
variables are anticommuting in a fixed column ($I=J$). The
$GL(D)$-irreducibility condition (ii) of Subsection
\ref{BeBo:genlingr} is implemented by the conditions
\begin{equation}
\Big(\,d_{_I}x\cdot\frac{\partial^L}{\partial (d_{_J}x)}\,-\,
\delta_{_{IJ}}\,\ell_{_I} \,\Big)\Phi\,=\,0\,,\quad(I\leqslant
J)\label{BeBo:GLD}
\end{equation}
where the dot stands for the contraction of the indices, $\ell_I$
for the length of the $I$th column in the Young diagram $\lambda$
and $\partial^L$ stands for ``left" derivative. By the Weyl
construction, an element $\Phi\in{\cal A}
$
satisfying (\ref{BeBo:GLD}) belongs to the Schur module
$V_\lambda^{GL(D)}\,$. Following the discussion of Subsection
\ref{BeBo:ortho}, if $\lambda$ denotes an allowed Young diagram, 
such an element $\Phi\in V_\lambda^{GL(D)}$ is
irreducible under the (pseudo)-orthogonal group $O(p,q)$ ($p+q=D$)
if it is traceless, that is
\begin{equation}
\Big(\,\frac{\partial^L}{\partial
(d_{_I}x)}\cdot\frac{\partial^L}{\partial
(d_{_J}x)}\,\Big)\Phi\,=\,0\,, \quad (\forall\; I,J)
\label{BeBo:OD}
\end{equation}
where the dot stands now for the contraction of indices via the
use of the metric preserved by $O(p,q)$. An element
$\Phi\in{\cal A}
$ such that
(\ref{BeBo:GLD})-(\ref{BeBo:OD}) are fulfilled belongs to the
Schur module $V_\lambda^{O(p,q)}$ labeled by the Young
diagram $\lambda\,$.

The Schur module $V_\lambda^{GL(D)}$ admits another convenient
realization in terms of polynomials in $r\times D$ commuting
variables. In other words, the vector space $V_\lambda^{GL(D)}$ is
isomorphic to a subspace of the polynomial algebra in the
variables $u^\mu_a$ ($a=1,2,\ldots,r$) where the index $a$
corresponds to each row. The condition (a) of Subsection
\ref{BeBo:genlingr} is automatically verified for any such
polynomial due to the fact that the variables are commuting in a
fixed row. The $GL(D)$-irreducibility condition (b) of Subsection
\ref{BeBo:genlingr} is implemented by the conditions
\begin{equation}
\Big(\,u_a\cdot\frac{\partial}{\partial u_b}\,-\,
\delta_{_{ab}}\,\lambda_a \,\Big)\Phi\,=\,0\,,\quad(a\leqslant
b)\label{BeBo:GLD'}
\end{equation}
where the dot still stands for the contraction of the indices. The
degree of homogeneity of the polynomial $\Phi$ in the variables
$u^\mu_a$ (for fixed $a$) is $\lambda_a\,$. The corresponding
coefficients are tensors irreducible under the general linear
group. By the Weyl construction, a polynomial $\Phi(u_a)$
satisfying (\ref{BeBo:GLD'}) belongs to the Schur module
$V_\lambda^{GL(D)}\,$. Again, such an element $\Phi\in
V_\lambda^{GL(D)}$ is irreducible under the (pseudo)-orthogonal
group $O(p,q)$ ($p+q=D$) iff it is traceless, that is
\begin{equation}
\Big(\,\frac{\partial}{\partial u_a}\cdot\frac{\partial}{\partial
u_b}\,\Big)\Phi\,=\,0\,, \quad (\forall\; a,b)\label{BeBo:OD'}
\end{equation}
where the dot stands for the contraction of indices via the use of
the metric preserved by $O(p,q)$. A polynomial $\Phi(u_a)$ such
that (\ref{BeBo:GLD'})-(\ref{BeBo:OD'}) are fulfilled belongs to
the Schur module $V_\lambda^{O(p,q)}$ labeled by an allowed Young
diagram $\lambda\,$.

\vspace{2mm}\noindent\textbf{Example:} Consider an irreducible
representation of the orthogonal group $O(D)$ labeled by the Young
diagram $\{n\}$ made of a single row of length equal to an integer
$n\,$. The polynomial $\Phi(u)\in V_{\{n\}}^{O(D)}$ obeys to the
irreducibility conditions
\begin{equation} \Big(\,u\cdot\frac{\partial}{\partial u}\,-\, n
\,\Big)\Phi\,=\,0\,,\quad\quad \Big(\,\frac{\partial}{\partial
u}\cdot\frac{\partial}{\partial u}\,\Big)\Phi\,=\,0\,.
\end{equation}
They mean that the polynomial is homogeneous (of degree equal to
$n$) and harmonic, so that its components correspond to a
symmetric traceless tensor of rank $n\,$:
$$\Phi(u)\,=\,\frac{1}{n!}\Phi_{\mu_1\ldots\mu_n}\,u^{\mu_1}\ldots u^{\mu_n}\,,
\quad\quad
\delta^{\mu_1\mu_2}\Phi_{\mu_1\mu_2\mu_3\ldots\mu_n}=0\,.$$ Of
course the integral of the square of such a polynomial over
${\mathbb R}^D$ is, in general, infinite. But the restriction of
an harmonic polynomial on the unit sphere $\overrightarrow{u}^2=1$
is square integrable on $S^{D-1}$. This restriction is called a
\textsl{spherical harmonic} of degree $n\,$. Therefore the space
of spherical harmonics of degree $n$ provides an equivalent
realization of the Schur module $V_{\{n\}}^{O(D)}\,$. For $D=3\,$,
the space $V_{\{n\}}^{O(3)}\,$ is spanned by the usual spherical
harmonics $Y^m_n(\theta,\phi)$ on the two-sphere with
$|m|\leqslant n\,$.

\vspace{2mm}\noindent\textbf{Remarks:}

\noindent$\bullet$ The infinitesimal generators of the
pseudo-orthogonal group $O(p,q)$ are represented by the operators
$$M_{\mu\nu}=i\sum\limits_{a=1}^{r}u_a^\rho\Big(g^{}_{\rho\mu}\frac{\partial}{\partial
u^\nu_a}-g^{}_{\rho\nu}\frac{\partial}{\partial u^\mu_a}\Big)\,.$$
Reordering the factors and making use of
(\ref{BeBo:GLD'})-(\ref{BeBo:OD'}) allows to reproduce the formula
(\ref{BeBo:quadrat}) for the eigenvalues of the quadratic Casimir
operator.

\noindent$\bullet$ Instead of polynomial functions in the
commuting variables, one may equivalently consider
\textit{distributions} obeying to the same conditions. The space
of solutions would carry an equivalent irrep, as follows from the
highest-weight construction of the representation. However, it
does not make sense any more of talking about the ``coefficients"
of the homogeneous distribution so that the link with the
equivalent tensorial representation is more intricate.

\vspace{2mm}The example of the spherical harmonics suggests that
it might be convenient to realize any unitary module of the
orthogonal group $O(D)$ as a space of functions on the unit
hypersphere $S^{D-1}$ satisfying some linear differential
equations. Better, the symmetry under the orthogonal group would
be made manifest by working with homogeneous harmonic functions on
the \textsl{ambient space} ${\mathbb R}^D\,$, evaluated on any
hypersphere $S^{D-1}\subset {\mathbb R}^D\,$.

\vspace{2mm}\noindent\textbf{Spherical harmonics:} \textit{To any
UIR of the isometry group $O(D)$ of a hypersphere $S^{D-1}$, one
may associate manifestly covariant differential equations for
functions on $S^{D-1}$ embedded in ${\mathbb R}^D$ whose space of
solutions carry the corresponding UIR.}\vspace{1mm}

\vspace{2mm}\noindent\underline{Proof:} Any UIR of the isometry
group $O(D)$ corresponds to a Schur module $V_\lambda^{O(D)}$
which may be realized as the space of polynomials
$\Phi(\overrightarrow{u}_a)$ such that
(\ref{BeBo:GLD'})-(\ref{BeBo:OD'}) are obeyed. 
Let us introduce the notation:
$\overrightarrow{x}:=\overrightarrow{u}_1$ and
$\overrightarrow{t}_{a-1}:=\overrightarrow{u}_{a}$ for
$a=2,\ldots,r\,$. One interprets the polynomial
$\Phi(\overrightarrow{x},\overrightarrow{t}_{\underline{a}})$
(where the index $\underline{a}$ runs from $1$ to $r-1$) as a
tensor field on the Euclidean space ${\mathbb R}^D$ parametrized
by the Cartesian coordinates $\overrightarrow{x}\,$, with some
auxiliary variables $\overrightarrow{t}_{\underline{a}}$ implementing
the tensor components. The conditions
(\ref{BeBo:GLD'})-(\ref{BeBo:OD'}) for $a$ and $b$ strictly
greater than $1$ imply that
\begin{equation}
\Big(\,t_{\underline{a}}\cdot\frac{\partial}{\partial
t_{\underline{b}}}\,-\,
\delta_{_{\underline{a}\underline{b}}}\,\lambda_{\underline{a}}
\,\Big)\Phi\,=\,0\,,\quad(\underline{a}\leqslant
\underline{b})\quad\quad \Big(\,\frac{\partial}{\partial
t_{\underline{a}}}\cdot\frac{\partial}{\partial
t_{\underline{b}}}\,\Big)\Phi\,=\,0\,,
\end{equation}
where $\underline{\lambda}=\{\lambda_2,\ldots,\lambda_r\}$ is the
Young diagram obtained from $\lambda$ by removing its first row.
Thus the components of the ``tensor field"
$\Phi(\overrightarrow{x},\overrightarrow{t}_{\underline{a}})$ carry an
irreducible representation of $O(D)$ labeled by
$\underline{\lambda}$. The conditions (\ref{BeBo:GLD'}) for
$a=b=1$ imply that
$$\Big(x\cdot\frac{\partial}{\partial x}-\lambda_1\Big)\Phi=0\,,$$
so the polynomial $\Phi(\overrightarrow{x},\overrightarrow{t}_{\underline{a}})$
is homogeneous of degree $\lambda_1$ in the radial coordinate
$|\overrightarrow{x}|\,$. 
The condition (\ref{BeBo:OD'}) for $a=b=1$ is interpreted as the 
Laplace equation
\begin{equation}\Big( \,\frac{\partial}{\partial x}
\cdot\frac{\partial}{\partial x}\,\Big)\Phi\,=\,0
\label{BeBo:laplace}
\end{equation} 
on the ambient space ${\mathbb R}^D$, it imples that the tensor field
$\Phi$ is harmonic in ambient space. The condition
(\ref{BeBo:GLD'}) for $b>a=1$ states that the radial components
vanish,
\begin{equation}
\Big(\,x\cdot\frac{\partial}{\partial t_{\underline{a}}} \,\Big)
\Phi\,=\,0\,,
\end{equation} 
so the tensor 
components are longitudinal to the hyperspheres $S^{D-1}\,$.
Therefore the evaluation of the non-vanishing components of
$\Phi(\overrightarrow{x},\overrightarrow{t}_{\underline{a}})$ on 
the unit hypersphere $|\overrightarrow{x}|=1$ is an \textsl{intrinsic}
tensor field living on the hypersphere $S^{D-1}$ and whose tensor 
components carry an irrep of the stability subgroup $O(D-1)$ labeled by
$\underline{\lambda}\,$. These tensor fields generalize the
spherical harmonics to the generic case $r\geqslant 1\,$. Finally, the condition (\ref{BeBo:OD'})
for $b>a=1$ states that the tensor field is divergenceless in
ambient space,
\begin{equation}\Big(\,\frac{\partial}{\partial
x}\cdot\frac{\partial}{\partial
t_{\underline{a}}}\,\Big)\Phi\,=\,0\,.\label{BeBo:divod}\end{equation}
The differential equations (\ref{BeBo:laplace}) and
(\ref{BeBo:divod}) are written in ambient space but they may be
reformulated in intrinsic terms on the hypersphere, at the price
of losing the manifest covariance under the full isometry group
$O(D)\,$.\qed

\subsection{Euclidean group}
\label{BeBo:IWcontraction}

The method of induced representations was introduced in Subsection
\ref{BeBo:inducedreps} for the Poincar\'e group
$ISO(D-1,1)^\uparrow$ and applied to the Euclidean group
$ISO(D-2)$ in Subsection \ref{BeBo:classification}. Focusing on
the faithful (\textit{i.e.} with a non-trivial action of the
translation generators) irreps of the \textit{inhomogeneous}
orthogonal group, all of them are induced from an UIR of the
stability subgroup. Using the results of the previous section
\ref{BeBo:ortho}, one may summarise the final result into the
following classification.\vspace{2mm}

\noindent{\textbf{Unitary irreps of the inhomogeneous orthogonal
groups:}} \textit{Each inequivalent UIR of the group $IO(D)$ with
a non-trivial action of its Abelian normal subgroup is associated
with a positive real number $\mu$ and an allowed Young diagram of the
subgroup $O(D-1)\,$, and conversely.}\vspace{2mm}

The orbits of the linear action of the orthogonal group $O(D)$ on
the Euclidean space ${\mathbb R}^D$ are the hyperspheres $S^{D-1}$
of radius $R\,$. The isometry group of any such hypersphere
$S^{D-1}$ is precisely $O(D)\,$. Considering a region of fixed
size on these hyperspheres, in the limit $R\rightarrow \infty$ the
sphere becomes a hyperplane ${\mathbb R}^{D-1}\,$. Therefore the
homogeneous and inhomogeneous orthogonal groups are related by
some infinite radius limit: $O(D)\rightarrow IO({D-1})$. Such a
process is frequently referred to as an
\textsl{In\"{o}n\"{u}-Wigner contraction} in the physics
literature \cite{BeBo:Inonu:1953sp}. This is better seen at the
level of the Lie algebra. Specializing the $D$th directions, the
commutation relations (\ref{BeBo:sodcomrel}) take the form
\begin{eqnarray}
i\,[M_{mn},M_{pq}] &=&
\delta_{np}M_{mq}-\delta_{mp}M_{nq}-\delta_{qm}M_{pn}
+\delta_{qn}M_{pm}\,,
\label{BeBo:1.7.10lca} \\
i\,[M_{mD},M_{pq}] &=& \delta_{mn}M_{pD}-\delta_{mp}M_{nD}\,,
\label{BeBo:1.7.10lcb} \\
i\,{[}M_{mD},M_{pD}{]} &=& M_{pm}\,. \label{BeBo:1.7.10cc}
\end{eqnarray}
where the latin letters take $D-1$ values. Defining
$M_{mD}\,=\,R\,P_m$ and taking the limit $R\rightarrow\infty$
(with $P_m$ fixed) in the relations
(\ref{BeBo:1.7.10lca})-(\ref{BeBo:1.7.10cc}) lead to
\begin{eqnarray}
i\,[M_{mn},M_{pq}] &=&
\delta_{np}M_{mq}-\delta_{mp}M_{nq}-\delta_{qm}M_{pn}
+\delta_{qn}M_{pm}\,,
\label{BeBo:1.7.11lca} \\
i\,[P_m,M_{pq}] &=& \delta_{mn}P_p-\delta_{mp}P_n\,,
\label{BeBo:1.7.11lcb} \\
i\,{[}P_m,P_{p}{]} &=& 0\,. \label{BeBo:1.7.11cc}
\end{eqnarray}
As can be seen, the generators $\{M_{mn},P_m\}$ span the Lie
algebra of the inhomogeneous orthogonal group $IO(D-1)\,$. The
former argument proves the contraction
$\mathfrak{so}(D)\rightarrow \mathfrak{iso}({D-1})\,$.

The limit of a sequence of irreps of the homogeneous orthogonal 
group $O(D)$,
in which one performs an In\"{o}n\"{u}-Wigner contraction, is
automatically a representation of the inhomogeneous orthogonal
group $IO(D-1)$ (if the limit is not singular). An interesting
issue is the inverse problem: which irreps of $IO(D-1)$ may be
obtained as the limit of such a sequence of irreps of $O(D)\,$?
The problem is non-trivial because, generically, the limit of a
sequence of irreps is a \textit{reducible} representation.
\vspace{2mm}

\noindent{\textbf{Contraction of UIRs of the homogeneous
orthogonal groups:}} \textit{Each inequivalent UIR of the group
$IO(D-1)$ with a non-trivial action of its Abelian normal subgroup
may be obtained as the contraction of a sequence of UIRs of the
group $O(D)$.}\vspace{2mm}

More precisely, the In\"{o}n\"{u}-Wigner contraction $R\rightarrow
\infty$ of a sequence of UIRs of $O(D)\,$, labeled by allowed
Young diagrams $\nu=\{s,\lambda_1,\ldots,\lambda_r\}$ such that
the limit of the quotient $s/R$ is a fixed positive real number $\mu$,
is the UIR of $IO(D-1)$ labeled by the parameter $\mu$ and the Young diagram
$\lambda=\{\lambda_1,\ldots,\lambda_r\}\,$.

\vspace{2mm}\noindent\underline{Proof:} The use of the spherical
harmonics construction discussed at the end of Subsection \ref{BeBo:auxiliary}
is very convenient here. The main idea is to solve the homogeneity condition in a neighborhood of $x^D\neq 0$
as follows:
\begin{equation}
\Phi(x^m,x^D,t_{\underline{a}})\,=\,z^s\,\phi\left(\frac{x^m}{z},t_{\underline{a}}\right)\,,
\label{homog}
\end{equation}
where $\overrightarrow{x}=(x^m,x^D)$ and $\phi(y^m,t_{\underline{a}}):=\Phi(y^m,\frac{s}{\mu},t_{\underline{a}})\,$.
In other words, one may perform a convenient change of coordinates from the \textsl{homogenous coordinates} $(x^m,x^D)$ to the
set $(y^m,z)$ where $$y^m=\frac{x^m}{z}$$ are the \textsl{inhomogenous coordinates} (on the projective space ${\mathbb P}{\mathbb R}^{D-1}$ minus the point at infinity $z=0\,$) and $$z=\frac{\mu\, x^D}{s}$$ is a scale variable.
The magic is that the equations for the generalized spherical harmonics have a well-behaved limit $x^D\rightarrow\infty$ in terms of $\phi(y^m,t_{\underline{a}})$ 
when $x^D/s$ is fixed to be equal to the ratio $z/\mu\,$, where $z$ and $\mu$ are finite \cite{BeBo:Bekaert:2005in}.
To see that, one should use the relations
\begin{eqnarray}
\frac{\partial}{\partial x^m}&=&\frac{1}{z}\,\frac{\partial}{\partial y^m}\,,\nonumber\\
\frac{\partial}{\partial x^D}&=&\frac{\mu}{s}\left(\frac{\partial}{\partial z}-
\frac{1}{z}\,y^m\frac{\partial}{\partial y^m}\right)\,.\label{rels}
\end{eqnarray}
Moreover, the equations in this limit may be identified with equations 
for the proper UIR of the inhomogeneous orthogonal
group $IO(D-1)$ realized homogeneously in terms of the inhomogenous coordinates.
\qed

\vspace{2mm}\noindent\textbf{Example:} The simplest instance is when $\lambda=\{0\}$ because one considers the
sequence of harmonic functions $\Phi(x^m,x^D)$ of homogeneity degree $s\,$. The Laplace operator acting on $\Phi(x^m,x^D)$ 
reads in terms of $\phi(y^m)$ as follows 
$$\Delta_{{\mathbb R}^D}\Phi=z^{s-2}\left[{\partial \over \partial y}\cdot{\partial \over \partial y}
+{\mu^2 \over s^2}\left(s(s-1)-(2s-1)\left(y\cdot{\partial \over
\partial y}\right)
+\left(y\cdot{\partial \over \partial y}\right)^2\right)\right]\phi\,,$$
due to the homogeneity condition (\ref{homog}) and the relations (\ref{rels}).
The Laplace equation $\Delta_{{\mathbb R}^D}\Phi=0$ is thus equivalent to the equation
$$\left[{\partial \over \partial y}\cdot{\partial \over \partial y}
+{\mu^2 \over s^2}\left(s(s-1)-(2s-1)\left(y\cdot{\partial \over
\partial y}\right)
+\left(y\cdot{\partial \over \partial y}\right)^2\right)\right]\phi=0\,,$$
whose limit for $s\rightarrow \infty$ is the Helmholtz equation
$[\,\Delta_{{\mathbb R}^{D-1}}+\mu^2\,]\,\phi=0\,,$ where $\Delta_{{\mathbb R}^{D-1}}={\partial \over \partial y}\cdot{\partial \over \partial y}\,$. The space of solutions of the Helmholtz equation
carries an UIR of $IO(D-1)$ induced from a trivial representation of the stability subgroup $O(D-2)\,$.

\section{Relativistic field equations}

The \textsl{Bargmann\,-Wigner programme} amounts to associating,
with any given UIR of the Poincar\'e group, a manifestly covariant
differential equation whose (positive-energy) solutions transform
according to the corresponding UIR. Physically, it might be
natural to restrict this programme to the two most important
classes of UIRs: the massive and massless representations.
Mathematically, this restriction is convenient because the
group-theoretical analysis is simpler since any of these UIRs is
induced from an UIR of a unimodular orthogonal group $SO(n)$ (with
$D-3\leqslant n\leqslant D-1$), as can be checked easily on the
tables of Subsection \ref{BeBo:classification}.

In 1948, this restricted programme was completed by Bargmann and
Wigner in four dimensions when, for each such UIR of
$ISO(3,1)^\uparrow\,$, a relativistic field equation was written
whose positive-energy solutions transform according to the
corresponding UIR \cite{BeBo:Bargmann48}. But this case ($D=4$)
will not be reviewed here in details because it may cast shadow on
the generic case. Indeed, it is rather peculiar in many respects:
\begin{itemize}
  \item The quadratic and quartic Casimir operators essentially classify
the UIRs, but this is no more true in higher dimensions where more
Casimir operators are necessary and the classification quickly
becomes technically cumbersome in this way. Moreover, one should
stress that the eigenvalues of the Casimir operators do not
characterize uniquely an irreducible representation (for instance,
the quadratic and quartic Casimir operators vanish for all
helicity representations).
  \item The (complex) Lorentz algebra
$\mathfrak{so}(3,1)$ is isomorphic to the direct sum of two
(complex) rotation algebras $\mathfrak{so}(3)\cong
\mathfrak{sp}(2)\,$. These isomorphisms allow the use of the
convenient ``dotted-undotted" formalism for the finite-dimensional
(non-unitary) irreps of the spin group $Spin(3,1)$.
  \item The symmetric tensor-spinor fields
are sufficient to cover all inequivalent cases.
 \item The helicity short little group $SO(2)$
is Abelian, therefore its irreps are one-dimensional, for fixed
helicity. Notice that the helicity is discretized because the
representation of the ``little group" $SO(2)$ is a restriction of
the representation of the group $Spin(3)\cong SU(2)$ which has no
intrinsically projective representations.
 \item The infinite-spin short little group $SO(1)$ is trivial,
 thus there are only two inequivalent infinite-spin representations
 (single- or double-valued) \cite{BeBo:Wigner39}.
 \item \textit{etc}.
\end{itemize}
Moreover, there exists an extensive literature on the subject of
UIRs of $ISO(3,1)^\uparrow$ and we refer to the numerous
pedagogical reviews available for more details on the
four-dimensional case (see \textit{e.g.} the inspiring
presentations of \cite{BeBo:Weinberg} and \cite{BeBo:Buchbinder}).

\vspace{2mm}It is standard to require time reversal and parity
symmetry of the field theory. More precisely, the field equations
we will consider are covariant under the two previous
transformations. As a consequence of the time reversal symmetry,
the representation is \textit{irreducible} under the group
$ISO(D-1,1)$ but \textit{reducible} under the Poincar\'e group
$ISO(D-1,1)^\uparrow\,$: the Hilbert space of solutions contain
both positive and negative energy solutions. Furthermore, the
parity symmetry implies that the representation is \textit{irreducible}
under the inhomogeneous Lorentz group $IO(D-1,1)$ but
\textit{reducible} under the group $ISO(D-1,1)$ (for instance,
both chiralities are present in the massless case for $D$ even). To conclude,
the Bargmann\,-Wigner programme is actually understood as
associating, with any given UIR of the inhomogeneous Lorentz
group, a manifestly covariant differential equation whose
solutions transform according to the corresponding UIR.

\subsection{General procedure}
\label{BeBo:strategy}
%

The lesson on induced representations that we learned from Wigner
implies the following strategy:
\begin{enumerate}
  \item Pick a unitary representation of the (short) little group.
  \item Introduce a wave function on ${\mathbb R}^{D-1,1}$
taking values in some (possibly non-unitary) representation of the
Lorentz group $O(D-1,1)$ the restriction of which to the (short)
little group contains the representation of step 1.
  \item Write a system of linear covariant equations, differential in position space $x^\mu$
thus algebraic in momentum space $p_\nu\,$, for the wave function
of step 2. These equations may not be independent.
  \item Fix the momentum and check in convenient coordinates
  that the field equations of step 3 put to zero all ``unphysical" components
  of the wave function. More precisely, verify that its non-vanishing
  components carry the unitary representation of step 1.
\end{enumerate}

\vspace{2mm}\noindent\underline{Proof:} The fact that the set of
linear differential equations is taken to be manifestly covariant
ensures that the Hilbert space of their solutions carries a 
(infinite-dimensional) representation of $IO(D-1,1)\,$. 
The fourth step determines the
representation of the little group by which it is induced. \qed

\vspace{1mm}In the physics literature, the fourth step is referred
to as ``looking at the physical degrees of freedom." If the
(possibly reducible) representation is proven to be unitary, then
this property is summarised in a ``no-ghost theorem."

\vspace{2mm}The Klein-Gordon equation $(p^2\pm\, m^2)\Psi=0$ is
always, either present in the system of covariant equations or a
consequence thereof. Consequently, the Klein-Gordon equation will
be assumed implicitly from now on in the step 3. Therefore, the
step 4 will be immediately performed in a proper Lorentz frame.
(We refer the reader to the Subsection \ref{BeBo:orbit} for more
details.)

The two completions \cite{Siegel:9912205} and \cite{BeBo:BBs} of the 
Bargmann\,-Wigner programme for finite-component representations 
in Minkowski spacetime of dimension 
$D>3$ are reviewed, respectively, in the appendix \ref{BeBo:SZeqs} 
and in the subsections \ref{BeBo:massive}-\ref{BeBo:masslessrep}
 for single-valued UIRs of the Poincar\'e group.\footnote{Spinorial irreps 
may be adressed analogously by supplementing the
system of differential equations with Dirac-like equations and
gamma-trace constraints (see \textit{e.g.}
\cite{BeBo:Bekaert:2005in,BeBo:Bandos:2005mb} for more details).}

The tachyonic case\footnote{The
discussion presented in the section \ref{BeBo:tachyonic} was not
published before, it directly derives from private conversations
between X.B. and J. Mourad.} is more
briefly discussed in Subsection \ref{BeBo:tachyonic}. 
The zero-momentum representations are 
not considered here since they essentially are the unitary irreducible 
representations of the de Sitter spacetime $dS_{D-1}\,$. The latter
have been reviewed in \cite{Basile:2016aen}.

The Bargmann\,-Wigner programme for fractional-spin fields in three spacetime dimensions 
has been completed in ~\cite{BeBo:Jackiw:1990ka}. More generally, the exhaustive 
completion of the Bargmann\,-Wigner programme
(for all representations) in Minkowski spacetime of dimension $D=3$ is briefly 
summarised in Appendix \ref{App:A}.

\subsection{Massive representations}
\label{BeBo:massive}
%

The Bargmann\,-Wigner programme is easy to complete for massive UIRs
because the massive stability subgroup is the orthogonal group
$O(D-1)\subset O(D-1,1)\,$. By going to a rest-frame, the
time-like momentum vector takes the form
$p^\mu=(m,\overrightarrow{0})\neq 0\,$. The physical components of
the field are thus carrying a tensorial irrep of the group
$O(D-1)$ of orthogonal transformations in the spatial hyperplane
${\mathbb R}^{D-1}\,$ orthogonal to $p^{\mu}\,$. 
In other words, the linear field equations
should remove all components including time-like directions. 
These unphysical components are responsible for the fact that the 
Fock space is not endowed with a positive-definite norm.

\textbf{Step 1.} From the sections \ref{BeBo:ABC} and
\ref{BeBo:Orthorepth}, one knows that any unitary representation
of the orthogonal group $O(D-1)$ is a sum of UIRs which are
finite-dimensional and thus, equivalent to a tensorial
representation. Let us consider the UIR of $O(D-1)$ labeled by the
allowed Young diagram $\lambda=
\{\lambda_1,\lambda_2,\ldots,\lambda_r\}$ (\textit{i.e.} the sum
of the lengths of its first two columns does not exceed $D-1$).

\textbf{Step 2.} The simplest\footnote{There are 
other possible equivalent representations. In the case $D=4\,$, see 
Sec. 5.7 of \cite{BeBo:Weinberg}, Eq. (5.7.33).} 
way to perform the Bargmann\,-Wigner
programme in the massive case is to choose a covariant wave
function whose components carry the (finite-dimensional and
non-unitary) tensorial irrep of the Lorentz group $O(D-1,1)$
labeled by the Young diagram $\lambda\,$. As explained in the
subsection \ref{BeBo:auxiliary}, a convenient way of realizing
this is in terms of a wave function $\Phi(p,u_a)$ polynomial in
the auxiliary commuting variables $u_a^\mu\,$ satisfying the
irreducibility conditions (\ref{BeBo:GLD'})-(\ref{BeBo:OD'}).

\textbf{Step 3.} The massive Klein-Gordon equation
\begin{equation}
(p^2+m^2)\Phi=0
\label{BeBo:KGequ}
\end{equation}
has to be
supplemented with the transversality conditions
\begin{equation}
\Big(\,p\cdot\frac{\partial}{\partial u_a}\,\Big)\Phi\,=\,0\,,
\label{BeBo:transvmass}
\end{equation}
of the wave function.

\textbf{Step 4.} Looking at a fixed-momentum mode in its
corresponding rest-frame $p^\mu=(m,\overrightarrow{0})$ leads to
the fact that the components of the wave function along the
timelike momentum are set to zero by (\ref{BeBo:transvmass}):
$\Phi=\Phi(p,\overrightarrow{u}_a)\,$. In words, $\Phi$ does not
depend on the time components $u^0_a\,$, $\forall\; a\,$.  
In this case, the conditions (\ref{BeBo:GLD'})-(\ref{BeBo:OD'}) 
read as irreducibility conditions under the orthogonal group 
$O(D-1)\,$. \qed

\vspace{2mm}\noindent\textbf{Example:} Massive symmetric
representations with ``spin" equal to $s$ correspond to Young
diagrams $\lambda=\{s\}$ made of one row of length equal to the
integer $s\,$. In four spacetime dimensions, this representation
is precisely what is usually called a ``massive spin-$s$
field."\footnote{To our knowledge, the Bargmann\,-Wigner programme
for the massive integer-spin representations in four-dimensional
Minkowski spacetime was adressed along the lines reviewed here for
the first time by Fierz in \cite{BeBo:Fierz1939}.} The covariant
wave function $\Phi(p,u)$ obeys to the irreducibility conditions
(\ref{BeBo:GLD'})-(\ref{BeBo:OD'}) of the components
\begin{equation}
\Big(\,u\cdot\frac{\partial}{\partial u}\,-\, s
\,\Big)\Phi\,=\,0\,,\quad\quad \Big(\,\frac{\partial}{\partial
u}\cdot\frac{\partial}{\partial u}\,\Big)\Phi\,=\,0\,.
\label{BeBo:masscalar}
\end{equation}
The wave function $\Phi$ is homogeneous of degree $s$ and harmonic
in the auxiliary variable $u\,$. If the wave function $\Phi(p,u)$
is polynomial in the auxiliary variable $u\,$, then its components
correspond to a symmetric tensor of rank $s$
$$\Phi(p,u)\,=\,\frac{1}{s!}\Phi_{\mu_1\ldots\mu_s}(p)\,u^{\mu_1}\ldots u^{\mu_s}\,,$$
which is traceless
\begin{equation}
\eta^{\mu_1\mu_2}\Phi_{\mu_1\mu_2\mu_3\ldots\mu_s}(p)=0\,.
\label{BeBo:tracelessym}
\end{equation}
The
covariant field equations are the massive Klein-Gordon equation
together with the transversality condition 
\begin{equation}
\Big(\,p\cdot\frac{\partial}{\partial u}\,\Big)\Phi\,=\,0\,,
\label{BeBo:transversalityequ}
\end{equation}
which reads in components as
\begin{equation}
p^{\mu_1}\Phi_{\mu_1\mu_2\ldots\mu_s}(p)=0\,.
\label{BeBo:transvmassivec}
\end{equation}
The non-vanishing components of a solution of
(\ref{BeBo:transvmassivec}) must be along the spatial directions,
\textit{i.e.} only $\Phi_{i_1\ldots i_s}(p)$ may be $\neq 0\,$.
This symmetric tensor field is traceless with respect to the
spatial metric: $\delta^{i_1i_2}\Phi_{i_1i_2i_3\ldots
i_s}(p)=0\,$, thus the physical components carry a symmetric irrep
of the orthogonal group $O(D-1)\,$, the dimension of which can be
computed by making use of the formula (\ref{BeBo:dimsym}). The
polynomial wave function $\Phi(p,u)$ evaluated on the internal
unit hypersphere $u^iu_i=1$ corresponds to a decomposition of the
physical components in terms of the spherical harmonics on the
internal hypersphere $S^{D-2}\,$, which is an equivalent, though
rather unusual, way of representing the physical components
(usually, the use of spherical harmonics is reserved to the
``orbital" part of the wave function).

\vspace{2mm}The quartic Casimir operator of the Poincar\'e algebra
is easily evaluated in components in the rest frame
\begin{eqnarray}
&-{1\over 2}&P^2M_{\mu\nu}M^{\mu\nu}\,\, +\,\,M_{\mu\rho}P^\rho\,
M^{\mu\sigma}P_\sigma\nonumber\\\quad &&=\,{1\over
2}\,\,m^2(M_{ij}M^{ij}+2M_{i0}M^{i0})\,-\,m^2M_{i0}M^{i0}
\,=\,m^2\,{1\over 2}\,M_{ij}M^{ij}\,,\nonumber \end{eqnarray}
giving as a final result for a massive representation associated
with a Young diagram $\lambda$
\begin{eqnarray}{\cal C}_4\Big(\mathfrak{iso}(D-1,1)\Big) &=& {\cal
C}_2\Big(\mathfrak{iso}(D-1,1)\Big)\,{\cal
C}_2\Big(\mathfrak{so}(D-1)\Big)\,\,, \nonumber\\
&=& m^2\, \sum_{a=1}^r\lambda_a(\lambda_a+D-2a-1)\,,
\label{BeBo:massiveC}
\end{eqnarray}
where the eigenvalues of the quadratic Casimir operator of the
rotation algebra are given by the formula (\ref{BeBo:quadrat}).

\vspace{2mm}\noindent\textbf{Example:} In any dimension $D\,$, the
eigenvalue of the quartic Casimir operator for a massive symmetric
representation of rank $s$ is equal to $m^2\,s(s+D-3)$. In four
spacetime dimensions, the square of the Pauli-Lubanski vector
acting on a massive field of spin-$s$ is indeed equal to
$m^2\,s(s+1)$. \vspace{1mm}

Each massive representation in $D\geqslant 4$ dimensions may
actually be obtained as the first Kaluza--Klein mode in a
dimensional reduction from $D+1$ down to $D$ dimensions. There is
no loss of generality because the massive little group $SO(D-1)$
in $D$ dimension is identified with the $(D+1)$-dimensional
helicity (short) little group. Such a Kaluza--Klein mechanism
leads to a St\"uckelberg formulation of the massive field.

The massless limit $m\rightarrow 0$ of a massive irrep with
$\lambda$ fixed is, in general, reducible because the irrep of the
massive little group $SO(D-1)$ is restricted to the helicity
(short) little group $SO(D-2)\subset SO(D-1)$. This argument
combined with the known branching rule for $O(D-1)\downarrow
O(D-2)$ (reviewed in Subsection \ref{BeBo:ortho}) allows to prove
that the massless limit of a massive irrep of the homogeneous
Lorentz group labeled by a fixed Young diagram $\lambda$ contains
each helicity irrep labeled by Young diagrams $\mu$ such that
$$\lambda_1\geqslant\mu_1\geqslant\lambda_2\geqslant\mu_2
\geqslant\ldots\geqslant\mu_{r-1}\geqslant\lambda_r\geqslant\mu_r\geqslant
0\,,
$$ with multiplicity one.
The zero modes of a dimensional reduction from $D+1$ down to $D$
dimensions are determined by the same rule.

\vspace{2mm}\noindent\textbf{Example:} The zero modes of the
dimensional reduction of a massive symmetric representations with
``spin" equal to $s$ are all helicity symmetric representations
with integer ``spins" not greater than the integer $s$, each with
multiplicity one. For the dimensional reduction of a gravitational
theory (\textit{i.e.} a spin-two particle), one recovers the usual
result that the massless spectrum is made of one ``graviton"
(spin-$2$), one ``photon" (spin-$1$) and one ``dilaton"
(spin-$0$).

\subsection{Massless representations}\label{BeBo:masslessrep}
%

The quartic Casimir operator of the Poincar\'e algebra is
evaluated easily in components in the light-cone coordinates (see
Subsection \ref{BeBo:orbit} for notations),
\begin{eqnarray}
-{1\over 2}\,P^2\,M_{\mu\nu}M^{\mu\nu}\,\,
+\,\,M_{\mu\rho}P^\rho\, M^{\mu\sigma}P_\sigma\,=\,0\,+\,M_{m +
}P^+\, M^{m-}P_-\,=\,\pi_m\pi^m \,,\nonumber
\end{eqnarray} giving as a final result for a massless
representation
\begin{equation}{\cal C}_4\Big(\mathfrak{iso}(D-1,1)\Big) \,=\, {\cal
C}_2\Big(\mathfrak{iso}(D-2)\Big)\, =\, \mu^2
\label{BeBo:masslessC4}
\end{equation}
where the quadratic Casimir operator of the massless little group
is written in (\ref{BeBo:quadratinfinitespin}).

\subsubsection{Helicity representations}\label{BeBo:helicityreps}
%

Helicity representations correspond to the case $\mu=0\,$, so that
$\pi^m=0$ and in practice the representation is induced from a
representation of the orthogonal group $O(D-2)\,$.

\vspace{1mm}\textbf{Step 1.} Again, any unitary representation of
the orthogonal group $O(D-2)$ is a sum of finite-dimensional UIRs.
Let us consider the UIR of the helicity short little group
$O(D-2)$ labeled by the allowed Young diagram $\lambda=
\{\lambda_1,\lambda_2,\ldots,\lambda_r\}$ (that is, the sum of the
lengths of its first two columns does not exceed $D-2$):
\begin{equation}
\begin{picture}(75,70)(-20,2)
\put(-25,30){$\lambda=$}
\multiframe(0,0)(13.5,0){1}(15,10){}\put(20,0){$\lambda_r$}
\multiframe(0,10.5)(13.5,0){1}(20,10){}\put(25,10.5){$\lambda_{r-1}$}
\multiframe(0,21)(13.5,0){1}(40,20){\ldots}\put(45,25){$\vdots$}
\multiframe(0,41.5)(13.5,0){1}(55,10){} \put(60,41.5){$\lambda_3$}
\multiframe(0,52)(13.5,0){1}(70,10){} \put(75,52){$\lambda_2$}
\multiframe(0,62.5)(13.5,0){1}(110,10){}
\put(115,62.5){$\lambda_1$}
\end{picture}\quad\quad.\label{BeBo:Young}
\end{equation}

The step 2 is more subtle to perform than for massive
representations because the field equations must set to zero all
components along the light-cone of the covariant wave function,
because they are unphysical. In other words, the covariant wave
equations should remove \textit{two} directions, and not only one
like in the massive case. This fact implies that the
transversality is not a sufficient condition any more, it must be
supplemented either by other equations or by gauge symmetries
asserting that one may quotient the solution space by pure gauge
fields. In these lecture notes, one focuses on two
\textit{gauge-invariant} formulations which may be respectively
referred to as ``Bargmann\,-Wigner formulation" in terms of the
field strength and ``gauge-fixed formulation" in terms of the
potential.

\vspace{3mm}\textit{Bargmann\,-Wigner equations}

\vspace{1mm}The so-called ``Bargmann\,-Wigner equations" were
actually first written by Dirac \cite{BeBo:Dirac} in
four-dimensional Minkowski spacetime in spinorial form. Their name
originates from their decisive use in the completion of the
Bargmann\,-Wigner programme \cite{BeBo:Bargmann48}. The
generalization of the Bargmann\,-Wigner equations to any dimension
was presented in \cite{BeBo:BBs} for tensorial irreps (reviewed
here) and in \cite{BeBo:Bandos:2005mb} for spinorial irreps. The latter programme had previously been completed in \cite{Siegel:1986zi} with different equations.

\vspace{1mm}\textbf{Step 2.} Let ${\overline{\lambda}}=
\{\lambda_1,\lambda_1,\lambda_2,\ldots,\lambda_r\}$ be the Young
diagram depicted as
\begin{equation}
\begin{picture}(75,80)(-20,2)
\put(-25,40){$\overline{\lambda}=$}
\multiframe(0,0)(13.5,0){1}(15,10){}\put(20,0){$\lambda_r$}
\multiframe(0,10.5)(13.5,0){1}(20,10){}\put(25,10.5){$\lambda_{r-1}$}
\multiframe(0,21)(13.5,0){1}(40,20){\ldots}\put(45,25){$\vdots$}
\multiframe(0,41.5)(13.5,0){1}(55,10){} \put(60,41.5){$\lambda_3$}
\multiframe(0,52)(13.5,0){1}(70,10){} \put(75,52){$\lambda_2$}
\multiframe(0,62.5)(13.5,0){1}(110,10){}\put(115,62.5){$\lambda_1$}
\multiframe(0,73)(13.5,0){1}(110,10){}\put(115,73){$\lambda_1$}
\end{picture}\quad\quad.\label{BeBo:Youngbar}
\end{equation}
It is obtained from the Young diagram $\lambda$ represented in
(\ref{BeBo:Young}) by adding a row of equal length on top of the
first row of $\lambda\,$. The Young diagram ${\overline{\lambda}}$
has at least two rows of equal lengths and the sum of the lengths of
its first two columns does not exceed $D\,$. The covariant wave
function is chosen to take values in the Schur module
$V_{\overline{\lambda}}^{O(D-1,1)}$ realized in the manifestly
antisymmetric convention. Following Subsection
\ref{BeBo:auxiliary}, the wave function ${\cal K}(p,d_{_I}x)$ is
taken to be a polynomial in the graded variables $d_{_I}x^\mu$
($\,I=1,2,\ldots,\lambda_1\,$) obeying the commutation relations
(\ref{BeBo:presented}). Moreover, the irreducibility conditions of
the components under the Lorentz group $O(D-1,1)$ are
\begin{equation}
\Big(\,d_{_I}x^\mu\frac{\partial^L}{\partial (d_{_J}x^\mu)}\,-\,
\delta_{_{IJ}}\,\overline{\ell}_{_I} \,\Big){\cal
K}\,=\,0\,,\quad(I\leqslant J)\label{BeBo:GLDK}
\end{equation}
where $\overline{\ell}_I$ stands for the length of the $I$th
column in the Young diagram $\overline{\lambda}\,$, and
\begin{equation}
\Big(\,\eta^{\mu\nu}\frac{\partial^L}{\partial
(d_{_I}x^\mu)}\frac{\partial^L}{\partial
(d_{_J}x^\nu)}\,\Big){\cal K}\,=\,0\,.
\label{BeBo:ODK}
\end{equation}

\textbf{Step 3.} The covariant field equations may be summarised in
the assertion that the wave function is a ``harmonic" multiform in
the sense that, $\forall\; I$, it is ``closed" 
\begin{equation} \Big(\,p_\mu
\,d_{_I}x^\mu\,\Big){\cal K}\,=\,0\,,
\label{BeBo:closed}
\end{equation}
and ``coclosed" (\textit{i.e.} transverse)
\begin{equation}
\Big(\,p^\mu\frac{\partial^L}{\partial (d_{_I}x^\mu)} \,\Big)
{\cal K}\,=\,0\,.
\label{BeBo:coclosed}
\end{equation}
The operators $p\cdot d_{_I}x$ act as ``exterior differentials"
(or ``curls"), they are nilpotent and obey graded commutation
relations. As one can easily see, the field equations
(\ref{BeBo:closed}) and (\ref{BeBo:coclosed}), considered together, 
imply the massless Klein-Gordon equation. Actually, the equations (\ref{BeBo:closed})
may even be imposed off-shell, whereas the equations (\ref{BeBo:coclosed})  
only hold on-shell~\cite{BeBo:BBs}.  

\textbf{Step 4.} In the light-cone frame (see Section
\ref{BeBo:Lorentzgr}), the components of the momentum may be taken
to be $p_{\mu}=(p_{-},0,0,\ldots, 0)$ with $p_-\neq 0\,$. On the
one hand, the transversality condition (\ref{BeBo:coclosed})
implies that the wave function does not depend on the variables
$d_{_I}x^+\,$. On the other hand, the closure condition
(\ref{BeBo:closed}) reads $(p_-d_{_I}x^-)\,{\cal K}=0\,$, the
general solution of which is ${\cal K}\,=\,(\prod_I
p_-d_{_I}x^-)\,\phi\,$, where $\phi$ depends neither on
$d_{_I}x^-$ nor on $d_{_I}x^+$ (due to the transversality
condition). In other words, the directions along the light-cone
have been removed, since $\phi=\phi(p,d_{_I}x^m)\,$ ($m=1,2,\ldots,D-2$). 
Focusing on this field, one may show that the irreducibility conditions
(\ref{BeBo:GLDK}) become, in terms of the function $\phi$, 
\begin{equation} \Big(\,d_{_I}x^m\frac{\partial^L}{\partial
(d_{_J}x^m)}\,-\, \delta_{_{IJ}}\,\ell_{_I}
\,\Big)\phi\,=\,0\,,\quad(I\leqslant J)\label{BeBo:GLDf}
\end{equation}
where $\ell_{_I}=\overline{\ell}_{_I}-1$, and the trace conditions
(\ref{BeBo:ODK}) implies
\begin{equation}
\Big(\,\delta^{mn}\frac{\partial^L}{\partial
(d_{_I}x^m)}\frac{\partial^L}{\partial
(d_{_J}x^n)}\,\Big)\phi\,=\,0\,.\label{BeBo:ODf}
\end{equation}
Since $\ell_{_I}$ is the length of the $I$th column of the Young
diagram $\lambda\,$, the system of equations
(\ref{BeBo:GLDf})-(\ref{BeBo:ODf}) states that the components of
the function $\phi$ carry a tensorial irrep of the orthogonal
group $O(D-2)\,$. Therefore, the same is true for the physical components
of the wave function ${\cal K}\,$.\qed

This may be reformulated covariantly by saying that the closure
(\ref{BeBo:closed}) of the wave function implies that
\begin{equation}
{\cal K}\,=\,\Big(\,\prod_{I=1}^{\lambda_1} p_\mu
d_{_I}x^\mu\,\Big)\,\phi\,. \label{BeBo:curl}\end{equation} In
components, this means that the tensor ${\cal K}$ is equal to
$\lambda_1$ curls of the tensor $\phi\,$. This motivates the name
``field strength" for the wave function ${\cal K}(p,d_{_I}x)\,$,
the components of which are irreducible under the Lorentz group
(when evaluated on zero\,-mass shell) and labeled by $\overline{\lambda}\,$, 
and the name ``potential" or ``gauge field" for the wave function 
$\phi(p,d_{_I}x)\,$, the
components of which may be taken to be irreducible under the
general linear group, with symmetries labeled by the
Young diagram $\lambda\,$.

\vspace{2mm}\noindent\textbf{Examples:}

\noindent$\bullet$ The helicity vectorial representation
corresponds to a Young diagram $\lambda=\{1\}$ made of a single
box. In four spacetime dimensions, this representation is
precisely what is usually called a ``vector gauge field".  The
Young diagram $\overline{\lambda}=\{1,1\}$ is a single column made
of two boxes. The wave function in momentum space is given by
$${\cal K}\,=\,\frac{1}{2}\,{\cal K}_{\mu\nu}(p) \,dx^{\mu}dx^{\nu}$$ 
which carries an irrep of $GL(D,\mathbb{R})$: 
the antisymmetric rank-two representation. 
As one can see, the wave function actually is a
differential two-form, the components of which transforming as an 
antisymmetric tensor of rank two. The field equations
(\ref{BeBo:closed}) and (\ref{BeBo:coclosed}), respectively, read in
components
$$p_{\mu}{\cal K}_{\nu\rho}+p_{\nu}{\cal K}_{\rho\mu}+p_{\rho}{\cal K}_{\mu\nu}=0
\quad{\rm{(Bianchi~identities)}}$$
and
$$p^{\mu}{\cal K}_{\mu\nu}=0\qquad {\rm{(transversality~conditions)}}\,.$$
The differential two-form $\cal K$ is indeed harmonic (closed and
coclosed). In physical terms, one says that the field strength
${\cal K}_{\mu\nu}$ obeys to the Maxwell equations. As usual, the
Bianchi identities imply that the field strength derives from a
potential: ${\cal K}_{\mu\nu}=p_\mu \phi_\nu-p_\nu \phi_\mu\,$. In
the light-cone coordinates, the transversality implies that the
components ${\cal K}_{+\nu}$ vanish, thus the only non-vanishing
components are ${\cal K}_{-n}=p_- \phi_n\,$. Therefore the only
physical components correspond to a $(D-2)$-vector in the hyperplane
transverse to the light-cone.

\noindent$\bullet$ Helicity symmetric representations with
``helicity" (or ``spin") equal to $s$ correspond to Young diagrams
$\lambda=\{s\}$ made of one row of length equal to the integer
$s\,$. In four spacetime dimensions, this representation is
precisely what is usually called a ``massless spin-$s$ field". 
The
Young diagram $\overline{\lambda}=\{s,s\}$ is a rectangle made of
two row of length equal to the integer $s\,$. The wave function is
thus a polynomial in the auxiliary variables
$${\cal K}\,=\,\frac{1}{2^s}\,{\cal K}_{\mu_1\nu_1\,\mid\ldots\,\mid\,\mu_s\nu_s}
\,d_1x^{\mu_1}d_1x^{\nu_1}\,\ldots\,d_sx^{\mu_s}d_sx^{\nu_s}$$
satisfying the irreducibility equations
(\ref{BeBo:GLDK})-(\ref{BeBo:ODK}) with $\ell_I=2\,$, 
$\forall ~I\in \{1,\ldots,s\}\,$.
The tensor ${\cal K}$ is, by
construction, antisymmetric in each of the $s$ sets of two indices
\begin{eqnarray}
{\cal K}_{\mu_1\nu_1\,\mid\ldots\,\mid\,\mu_s\nu_s}\,=\,-\,{\cal
K}_{\nu_1\mu_1\,\mid\ldots\,\mid\,\mu_s\nu_s}=\,\ldots\,=\,
-\,{\cal K}_{\mu_1\nu_1\,\mid\ldots\,\mid\,\nu_s\mu_s}\,.
\end{eqnarray}
Moreover, the complete antisymmetrization over any set of three
indices gives zero and all its traces are zero on-shell, so that the 
on-shell tensor ${\cal K}$
indeed belongs to the space irreducible under the Lorentz group
$O(D-1,1)$ characterized by a two-row rectangular Young diagram of
length $s\,$. In four-dimensional Minkowski spacetime, the irrep
of the Lorentz group $O(3,1)$ carried by the on-shell  tensor ${\cal K}$ 
is usually denoted as $(s,0)\oplus(0,s)\,$.  More precisely, the
symmetry properties of the tensor 
${\cal K}_{\mu_1\nu_1\,\mid\ldots\,\mid\,\mu_s\nu_s}$ are labeled by the
Young tableau
\vspace{.1cm}

\begin{eqnarray}
\begin{picture}(83,10)(0,5)
\multiframe(1,14)(12.5,0){2}(12,12){\small $\mu_1$}{\small $\mu_2$}
\multiframe(25.5,14)(12.5,0){1}(42,12){$\ldots$}
\multiframe(68,14)(12.5,0){1}(12,12){\small $\mu_s$}
\multiframe(1,1.5)(12.5,0){2}(12,12){\small $\nu_1$}{\small
$\nu_2$} \multiframe(25.5,1.5)(12.5,0){1}(42,12){$\ldots$}
\multiframe(68,1.5)(12.5,0){1}(12,12){\small $\nu_s$}
\put(90,14){.}
\end{picture}\label{BeBo:Youngtabl}
\nonumber
\end{eqnarray}
The equation (\ref{BeBo:curl}) means that the components of the
tensor ${\cal K}_{\mu_1\nu_1\,\mid\ldots\,\mid\,\mu_s\nu_s}$ are
essentially the projection of 
$p_{\mu_1}\ldots p_{\mu_s}\phi_{\nu_1\ldots\nu_s}$ on the tensor 
field irreducible under $GL(D,\mathbb R)$ with symmetries labeled 
by the above Young tableau. The physical components
$\phi_{n_1\ldots n_s}$ of the symmetric tensor gauge potential
$\phi_{\nu_1\ldots\nu_s}$ are along the $D-2$ directions
transverse to the light-cone. The number of physical degrees of
freedom of a helicity symmetric field of rank $s$ can be computed
by making use of the formula (\ref{BeBo:dimsym}). 

\noindent $\bullet$ The helicity symmetric representation with
``spin" equal to $2$ corresponds to the graviton. The field strength
has the symmetry properties of the Riemann tensor. Its on-shell 
tracelessness indicates that it corresponds to the 
(linearized) Weyl tensor.
The equations (\ref{BeBo:closed}) are the Bianchi identities for 
the linearized Riemann tensor in flat spacetime, whereas 
the equations (\ref{BeBo:coclosed}) hold as a consequence of the 
sourceless Einstein equations linearized around flat spacetime. 
 
\vspace{2mm}\noindent\textbf{Remark:}
 One can find some early indications for the 
existence of the tensor ${\cal K}^{\mu_1\nu_1\,\mid\ldots\,\mid\,\mu_s\nu_s}$ 
in the paper \cite{BeBo:Weinberg:1965rz} 
where Weinberg constructs free quantum field operators that have a nonzero 
expectation value between the vacuum and one-particule states for 
massless particles of helicity $\pm s$ in four spacetime dimensions. 
In Weinberg's approach, one cannot find the classical (or ``first-quantized'') 
field strength tensor ${\cal K}^{\mu_1\nu_1\,\mid\ldots\,\mid\,\mu_s\nu_s}$
that we have built above, but instead a quantum \emph{operator} 
(in so-called ``second-quantization'') that we denote here 
$\widehat{\cal K}_{\pm}{}^{\mu_1\nu_1\,\mid\ldots\,\mid\,\mu_s\nu_s}$
and that transforms 
like a tensor under Lorentz transformations. This operator is built out of the product 
$[p^{\mu_{1}}e_{\pm}^{\nu_{1}}(\vec{p}\,)-p^{\nu_{1}}e_{\pm}^{\mu_{1}}(\vec{p}\,)]$
$\ldots $ $[p^{\mu_{s}}e_{\pm}^{\nu_{s}}(\vec{p}\,)-p^{\nu_{s}}e_{\pm}^{\mu_{s}}(\vec{p}\,)]$
featuring the two polarisation ``vectors'' $e_{\pm}^{\mu}(\vec{p\,})\,$. 
On the one hand, solving the Bianchi identities for the field strength 
${\cal K}^{\mu_1\nu_1\,\mid\ldots\,\mid\,\mu_s\nu_s}\,$ allows to write the latter  
as an expression involving $s$ derivatives of a completely symmetric 
gauge potential $\phi_{\mu_{1}\ldots\mu_{s}}\,$. 
This potential satisfies \cite{BeBo:BBs} the second-order Fronsdal field 
equations \cite{Fronsdal:1978rb} 
and is the building block for the construction of an interacting quantum 
field theory with long-range interactions.
On the other hand, the canonical quantization of the free field theory with field strength 
tensor ${\cal K}$ gives rise to Weinberg's quantum field operator $\widehat{\cal K}_{\pm}\,$.
The same remarks apply to the relation between the generalised field strength  
\eqref{BeBo:curl} and its second-quantized version in \cite{Weinberg:2020nsn}.

\vspace{3mm}\textit{Gauge-fixed equations}

The following equations are somewhat unusual, but they proved to
be crucial in the completion of the Bargmann\,-Wigner programme
for the infinite spin representations \cite{BeBo:Bekaert:2005in}.

\vspace{1mm}\textbf{Step 2.} Let $\widehat{\lambda}=
\{\lambda_1-1,\lambda_2-1,\ldots,\lambda_r-1\}$ be the Young
diagram depicted as
\begin{equation}
\begin{picture}(75,70)(-20,2)
\put(-25,30){$\widehat{\lambda}=$}
\multiframe(0,0)(13.5,0){1}(9,10){}\put(14,0){$\lambda_r-1$}
\multiframe(0,10.5)(13.5,0){1}(14,10){}\put(19,10.5){$\lambda_{r-1}-1$}
\multiframe(0,21)(13.5,0){1}(30,20){\ldots}\put(35,25){$\vdots$}
\multiframe(0,41.5)(13.5,0){1}(45,10){}
\put(50,41.5){$\lambda_3-1$} \multiframe(0,52)(13.5,0){1}(60,10){}
\put(65,52){$\lambda_2-1$}
\multiframe(0,62.5)(13.5,0){1}(100,10){}
\put(105,62.5){$\lambda_1-1$}
\end{picture}\quad\quad,\label{BeBo:Younghat}
\end{equation}
obtained from the Young diagram $\lambda$ represented in
(\ref{BeBo:Young}) by removing the first column of $\lambda\,$.
Therefore the sum of the length of the first two columns of the
Young diagram ${\widehat{\lambda}}$ does not exceed $D-2\,$. The
covariant wave function is chosen to take values in the Schur
module $V_{\widehat{\lambda}}^{O(D-1,1)}$ realized in the
manifestly symmetric convention. Actually, as anticipated in 
Subsection \ref{BeBo:auxiliary}, it turns out to be crucial to 
regard the wave function $\Phi(p,u_a)$ as a \textit{distribution} 
in the commuting auxiliary variables $u^\mu_a\,$, obeying to
\begin{eqnarray}
\left[\left(u_a\cdot{\partial \over \partial
u_b}\right)-\widehat{\lambda}_a\,\delta_{ab}\right]\Phi&=&0\,,\quad
(a\leqslant b)\,.\label{BeBo:gl}\\ \left({\partial \over
\partial u_a}\cdot{\partial \over \partial
u_b}\right)\Phi&=&0\,,\label{BeBo:o}
\end{eqnarray}

\vspace{1mm}\textbf{Step 3.} Proper field equations are the
transversality condition (\ref{BeBo:transvmass}) combined with the
equation
\begin{equation}
(p \cdot u_a)\,\Phi=0\,.\label{BeBo:lcone1}
\end{equation}
The equations (\ref{BeBo:lcone1}) and (\ref{BeBo:transvmass}) are
the respective analogues of the closure and coclosure conditions
(\ref{BeBo:closed})-(\ref{BeBo:coclosed}). A drastic difference is
that the operators $p\cdot u_a$ are not nilpotent (thus there is
no underlying cohomology). Actually, the equation
(\ref{BeBo:lcone1}) has no solution if $\Phi$ is assumed to be a 
polynomial in all the variables.

\vspace{1mm}\textbf{Step 4.} Equation (\ref{BeBo:lcone1}) can be
solved as 
\begin{equation}
\Phi=\delta(u_a\cdot p)
\,\phi\,,\label{BeBo:Phi}
\end{equation} 
where the distribution $\phi(p,u_a)$ may actually be assumed to 
be a function depending polynomially on the auxiliary variables 
$u_a$ for the present purpose. The Dirac delta is a distribution 
of homogeneity degree equal to minus one, hence the irreducibility 
conditions
(\ref{BeBo:gl})-(\ref{BeBo:o}) imply that
\begin{eqnarray}
\left[\left(u_a\cdot{\partial \over \partial
u_b}\right)-\lambda_a\,\delta_{ab}\right]\phi&=&0 \qquad
(a\leqslant b)\,,\label{BeBo:gl'}\\ \left({\partial \over
\partial u_a}\cdot{\partial \over \partial
u_b}\right)\phi&=&0\,.\label{BeBo:o'}
\end{eqnarray}
The function $\phi$ is defined from (\ref{BeBo:Phi}) modulo the
equivalence relation
\begin{equation}\phi\sim\phi+\sum\limits_{a=1}^r\,(u_a\cdot p)
\,\epsilon_a\label{BeBo:gsym}\end{equation} where $\epsilon_a$ are
arbitrary functions. This means that (\ref{BeBo:Phi}) is
equivalent to the alternative road towards the Bargmann\,-Wigner
programme: the gauge symmetry principle with the irreducible
components of $(u_a\cdot p) \,\epsilon_a$ being pure gauge fields.
As mentioned before, this path will not be addressed here (see
\textit{e.g.} \cite{BeBo:BBs} and refs therein for more
discussions on the gauge-invariance issue). Therefore, one may say that the
equation (\ref{BeBo:lcone1}) is the ``remnant" of the gauge
symmetries (\ref{BeBo:gsym}). In the light-cone coordinates, the
gauge symmetries (\ref{BeBo:gsym}) imply that one may choose a
representative $\phi$ which does not depend on the variables
$u^-_a$ (the gauge is ``fixed"). The transversality condition
(\ref{BeBo:transvmass}) implies that $\phi$ is also transverse, 
implying no dependence on $u^+_a$ (``gauge shoots twice"). 
Thus $\phi$ depends only on the
transverse auxiliary variables $u^m_a\,$, so one concludes by
observing that the physical components of $\phi$ carry a tensorial
irrep of $O(D-2)$ labeled by $\lambda\,$.\qed

\subsubsection{Infinite spin representations}\label{BeBo:infspin}
%

Infinite spin representations correspond to the case $\mu\neq 0$
and, in practice, the representation of the massless little group
$IO(D-2)$ is induced from a representation of the orthogonal group
$O(D-3)\,$. The parameter $\mu$ is a real parameter with the
dimension of a mass. Wigner proposed a set of manifestly covariant
equations to describe fields carrying these UIR in four spacetime
dimensions \cite{wi}. They have been generalized to arbitrary
infinite-spin representations in any 
dimension\cite{BeBo:Bekaert:2005in}.\footnote{More recent 
developments (as well as a list of open challenges) are reviewed in~\cite{Bekaert:2017}.}

\vspace{1mm}\textbf{Step 1.} Again, any unitary representation of
the orthogonal group $O(D-3)$ is a sum of finite-dimensional UIRs.
Let us consider the UIR of the helicity short little group
$O(D-3)$ labeled by the allowed Young diagram $\lambda=
\{\lambda_1,\lambda_2,\ldots,\lambda_r\}$ (that is, the sum of the
lengths of its first two columns does not exceed $D-3$).

\vspace{1mm}\textbf{Step 2.} In order to have manifest covariance,
it is necessary to lift the eigenvalues $\xi^m$ of the generators
$\pi^m$ in the massless little group to a $D$-vector $\xi^\mu\,$.
In practice, the covariant wave function is taken to be a distribution
$\Phi(p,\xi,u_a)\,$ satisfying the conditions
(\ref{BeBo:GLD'})-(\ref{BeBo:OD'}). The tensorial
components associated with the commuting variables $u_a$ belong to
the Schur module of the Lorentz group $O(D-1,1)$ labeled by an
allowed Young diagram $\lambda\,$.

\vspace{1mm}\textbf{Step 3.} Relativistic equations describing a
first-quantized particle with infinite spin are
\begin{eqnarray}
(p\cdot\xi)\,\Phi&=&0\,,\label{BeBo:epsilon3}\\
\left(p\cdot\frac{\partial}{\partial \xi}
-i\right)\Phi&=&0\,,\label{BeBo:epsilon1}\\
(\xi^2-\mu^2)\,\Phi&=&0\,, \label{BeBo:epsilon2}
\end{eqnarray}
together with the transversality conditions
\begin{eqnarray}(p\cdot  u_a)\,
\Phi&=&0\,,\label{BeBo:new2.5}\\\left(p\cdot {\partial \over
\partial u_a}\right)
\Phi&=&0\,,\label{BeBo:new3}\\
\left(\xi\cdot {\partial \over \partial u_a}\right) \Phi&=&0\,.
\label{BeBo:new4}
\end{eqnarray}
This system of equations is far from being independent. For
instance, compatibility condition of the systems
(\ref{BeBo:epsilon3})-(\ref{BeBo:epsilon1}) or
(\ref{BeBo:new2.5})-(\ref{BeBo:new3}) is the massless Klein-Gordon
equation.

\vspace{1mm}\textbf{Step 4.} The equation (\ref{BeBo:epsilon1})
reflects the fact that the couples $(p\,,\xi)$ and
$(p\,,\xi+\alpha p)$ are physically equivalent for arbitrary
$\alpha\in\mathbb R$. Indeed, one gets
\begin{equation}\Phi(p\,,\xi+\alpha p)\,=\,e^{i\alpha}\,
\Phi(p\,,\xi) \label{BeBo:jau} \end{equation} from Equation
(\ref{BeBo:epsilon1}). The equation (\ref{BeBo:epsilon2}) states
that the internal vector $\xi$ is a space-like vector while the
mass-shell condition states that the momentum is light-like. From
the equation (\ref{BeBo:epsilon3}), one obtains that the internal
vector is transverse to the momentum. All together, one finds that
$\xi$ may be taken to live on the hypersphere $S^{D-3}$ of radius
$\mu$ embedded in the transverse hyperplane ${\mathbb R}^{D-2}\,$.
In brief, the ``continuous spin" degrees of freedom essentially
correspond to $D-3$ angular variables, whose Fourier conjugates
are discrete variables analogous to the usual spin degrees of
freedom. Finally, proceeding analogously to the ``gauge-fixed"
field equations of the helicity representations, one may show 
\cite{BeBo:Bekaert:2005in} that
the conditions (\ref{BeBo:new2.5})-(\ref{BeBo:new4}) concretely
remove three unphysical directions in the components, so that the
final result is a tensorial irrep of the short little group
$O(D-3)$ fixing both the momentum $p$ and the internal vector
$\xi\,$.

\vspace{2mm}{}From the group theoretical point of view, the UIR of
the homogeneous and inhomogeneous orthogonal groups are related by
an In\"{o}n\"{u}-Wigner contraction $O(D-1)\rightarrow IO({D-2})$
(see Subsection \ref{BeBo:IWcontraction}). It follows that one can
obtain the continuous spin representations from the massive ones
in a suitable massless limit $m\rightarrow 0$ since their little
group UIRs are related by a contraction. The quartic Casimir
operator of the Poincar\'e group for the massive representation is
related to its Young diagram $\nu$ labeling the UIR of the little
group $O(D-1)$ via the formula (\ref{BeBo:massiveC}):
\begin{equation}{\cal C}_4\Big(\mathfrak{iso}(D-1,1)\Big)
= m^2\, \sum_{a=1}^r\nu_a(\nu_a+D-2a-1)\,, \label{BeBo:massiveC'}
\end{equation}
In order to keep ${\cal C}_4$ non-vanishing, the massless limit
must be such that the product of the ``spin" $\nu_1=s$ and the
mass $m$ remains finite. More precisely, one needs $sm\rightarrow
\mu\,$ in order to reproduce (\ref{BeBo:masslessC4}), so that the
spin goes to infinity while the row lengths $\nu_a$ for $a\neq
1\,$ are kept equal to $\lambda_{a-1}\,$
\cite{BeBo:Khan:2004nj,BeBo:Bekaert:2005in}. The Fourier transform
(in the internal space spanned by $\xi$) of the field equations
(\ref{BeBo:epsilon3})-(\ref{BeBo:new4}) may be obtained in this
way from the field equations of a massive representation in
``gauge-fixed" form (see \cite{BeBo:Bekaert:2005in} for more
details). This limit is very similar to the contraction of
Subsection \ref{BeBo:IWcontraction}.

\subsection{Tachyonic representations}
\label{BeBo:tachyonic}
%

The tachyonic representations have some similarities with the
massive representations. The simpler one is the analogue of the
Klein-Gordon equation, up to a change of sign for the mass term. The other
similarity is that the linear equations should remove the
components along the momentum. Of course, the major difference is
that the momentum is space-like. The quartic Casimir operator of the Poincar\'e algebra
is also evaluated easily in components,
giving as a final result for a tachyonic representation,
\begin{eqnarray}{\cal C}_4\Big(\mathfrak{iso}(D-1,1)\Big) &=& {\cal
C}_2\Big(\mathfrak{iso}(D-1,1)\Big)\,{\cal
C}_2\Big(\mathfrak{so}(D-2,1)\Big)\,\,, 
\label{BeBo:tachyomiC}
\end{eqnarray}
where the eigenvalues of the quadratic Casimir operator of the
rotation algebra are given by the formula (\ref{BeBo:quadrat}).

\textbf{Step 1.} The first step is more involved for the tachyonic
case since it requires the exhaustive knowledge of the UIR theory
for the groups $SO(D-2,1)^\uparrow\,$. Fortunately, complete
results are available \cite{BeBo:Bargmann,BeBo:Thieleker}. The steps
2-3 further require the completion of the Bargmann\,-Wigner
programme for the isometry group $SO(D-2,1)^\uparrow$ of the de
Sitter spacetime $dS_{D-2}\,$. This has been done in 
\cite{Basile:2016aen}.\footnote{The Bargmann\,-Wigner programme
in $AdS_{D}\,$, with field equations  that generalise the ones 
presented in Section \ref{BeBo:masslessrep}, were obtained 
in \cite{Boulanger:2008up}. 
Similar equations were obtained later in the $dS_{D}$ signature \cite{Basile:2016aen}.}

Let us assume that this programme has been performed through an
ambient space formulation, analogous to the one of the spherical
harmonics, as discussed in the subsection \ref{BeBo:auxiliary}.
More explicitly, let us consider that the physical components of
the wave function have been realized via a function on the
hyperboloid $dS_{D-2}$ of radius $\mu>0$ embedded in ${\mathbb
R}^{D-2,1}$ with some set of auxiliary commuting vectors of
${\mathbb R}^{D-2,1}$ (for the spin degrees of freedom) and the
corresponding $O(D-2,1)$-covariant field equations of the UIR are
known explicitly. The step 1 is therefore assumed to be performed.

\textbf{Step 2.} In order to have manifest Lorentz invariance, all
auxiliary variables are lifted to $D$-vectors: the coordinates of
the internal de Sitter spacetime are denoted by $\xi^\mu$ and the
auxiliary variables by $u_A^\mu\,$. The wave function is taken to
be $\Phi(p,\xi,u_A)\,$, where the internal vector $\xi$ plays a
role similar to the one in the infinite-spin representations. An
important distinction is that in the ambient space formulation,
one would evaluate the wave function on the hypersurface $\xi^2=\mu^2$
instead of imposing this relation on the wave function, as in
(\ref{BeBo:epsilon2}). The $O(D-2,1)$-covariant field equations for
the UIR of the little group $O(D-2,1)$ must be
$O(D-1,1)$-covariantized accordingly. Concretely, this implies
that the components of the covariant wave function carry an
(infinite-dimensional) irrep of the Lorentz group.

\textbf{Step 3.} These covariantized field equations and the
tachyonic Klein-Gordon equation $(p^2-m^2)\psi=0$ must be
supplemented by two equations: say the orthogonality condition
(\ref{BeBo:epsilon3}), similarly to the infinite spin
representation, and the transversality condition
(\ref{BeBo:transvmass}), similarly to the massive representation.
The orthogonality condition
(\ref{BeBo:epsilon3}) may be replaced by another transversality equation for the 
vector $\xi\,$.

\textbf{Step 4.} Now, the equation (\ref{BeBo:epsilon3}) implies
that the internal vector belongs to the hyperplane ${\mathbb
R}^{D-2,1}$ orthogonal to the momentum $p\,$. Its intersection
with the hypersurface $\xi^2=\mu^2$ restricts $\xi$ to the
internal de Sitter space $dS_{D-2}\subset{\mathbb R}^{D-2,1}$.
Moreover, the condition (\ref{BeBo:transvmass}) sets to zero all
components of the wave function along the momentum. Therefore, the
remaining components are physical and carry an UIR of the little
group $O(D-2,1)$ by construction (see step 2). \qed

\vspace{2mm}\noindent\textbf{Example:} The simplest non-trivial
example corresponds to a tachyonic representation of the
inhomogeneous Lorentz group $IO(D-1,1)$ induced by a
representation of the little group $O(D-2,1)$ corresponding to
``massive scalar field" on the ``internal de Sitter spacetime"
$dS_{D-2}$ with $D\geqslant 4\,$. This UIR belongs to the \textsl{principal continuous
series} of UIR of the group $O(D-2,1)$ and it may be realized as the space of
harmonic functions on ${\mathbb R}^{D-2,1}$ of (complex) homogeneity degree $s$ equal
to $\frac{3-D}{2}+i\,\sigma$ (with $\sigma$ a positive real parameter)
evaluated on the unit one-sheeted hyperboloid $dS_{D-2}\subset{\mathbb R}^{D-2,1}\,$. 
They can be regarded as a generalization of the spherical harmonics
in the Lorentzian case, where the degree is a complex number.
The eigenvalue of the quadratic Casimir operator (\ref{BeBo:QuadratiC}) 
of the little group $O(D-2,1)$ on this representation
is equal to
\begin{eqnarray}{\cal C}_2\Big(\mathfrak{so}(D-2,1)\Big)\,=
\,\left(\frac{D-3}{2}\right)^2+\sigma^2\,.
\end{eqnarray}
The d'Alembertian on the unit hyperboloid evaluated on such functions
is precisely equal to the former eigenvalue 
(as is true for the Laplacian on the unit sphere evaluated on spherical harmonics)
so the corresponding fields on the internal spacetime $dS_{D-2}$ are indeed ``massive". 
Inserting the above result in (\ref{BeBo:tachyomiC}), one sees that the quartic Casimir 
operator is negative for the corresponding tachyonic 
representation. In four-dimensional Minkowski spacetime, this implies that the 
Pauli-Lubanski vector is time-like.
The Lorentz-covariant wave function is taken to be $\Phi(p,\xi)$ evaluated on 
$\xi^2=1$ and the corresponding relativistic equations for the induced tachyonic 
representation may be chosen as 
\begin{eqnarray}
\left(p^2-m^2\right)\,\Phi&=&0\,,\label{BeBo:tachyo1}\\
\left(p\cdot\frac{\partial}{\partial \xi}\right)\,\Phi&=&0\,,\\
\left(\frac{\partial}{\partial \xi}\cdot\frac{\partial}{\partial \xi}\right)\Phi&=&0\,,\\
\left(\xi\cdot\frac{\partial}{\partial \xi}-s\right)\Phi&=&0\,,\label{homogen}
\end{eqnarray}
where one should remember that $s=\frac{3-D}{2}+i\,\sigma\,$. 
Notice the formal analogy with the system of equations 
(\ref{BeBo:KGequ}), \eqref{BeBo:transversalityequ} and 
(\ref{BeBo:masscalar}) for a massive symmetric tensor field.

\vspace{2mm}\noindent\textbf{Remark:} There might be sometimes
confusion in the folklore surrounding the tachyons. We would like
to insist on the fact that the tachyonic representations are
indeed \textit{unitary} (by definition). Still, their physical
interpretation is problematic because they are \textit{not causal}
in the sense that one may show that the support of their
propagator requires superluminal propagation. Roughly speaking,
the acausality is obvious because the momentum is space-like,
$p^2=+m^2\,$. The confusing point is that one may try to
circumvent this problem in the following way: solving $p^2-m^2=0$
by $p^\mu=(im,\overrightarrow{0})$ enforces causality, but the
price to pay is the loss of unitarity. Indeed, the energy is pure
imaginary, hence a naive plane-wave $e^{\pm i\,p_{_0}\,x^0}$ is
actually a non-integrable exponential $e^{\pm m x^0}\,$. These
remarks are summarised in the following table:

\begin{center}
\begin{tabular}{|c|c|c|c|}
  \hline
  $E=p_0$ & $|\overrightarrow{p}|$ & Unitarity & Causality \\\hline
  $0$ & $m$ & OK & KO \\
  $\pm i m$ & $0$ & KO & OK \\ \hline
\end{tabular}
\end{center}

Nevertheless, the tachyonic representations should not be
discarded too quickly on such physical grounds. Actually, if
tachyonic representations appear in the spectrum of a theory, then
it merely signals a local instability of the field theory in the
sense that the perturbation theory is performed around an unstable
vacuum, and the tachyon might roll to a stable vacuum (if any). For
instance, the Higgs particle is described by nothing but a
tachyonic scalar field (induced by the trivial representation of
the little group). By analogy, one may wonder if some
infinite-component tachyonic field (induced by a non-trivial
representation of the little group) could not play a role in some
huge Brout--Englert--Higgs mechanism providing mass to an infinite
tower of gauge fields in various massless irreps.

\section*{Acknowledgements}

We thank all the colleagues with whom we have worked and discussed on the 
Bargmann\,-Wigner programme in maximally symmetric spacetimes. 
The authors also acknowledge the Institut des Hautes \'Etudes
Scientifiques (Bures-sur-Yvette, France) and the Universit\'e de
Mons-Hainaut (Mons, Belgium) for hospitality during the first stage of this work.
The work of NB was partially supported by an F.R.S.-FNRS PDR grant 
``Fundamental issues in extended gravity'' No T.0022.19.

\begin{appendix}

\section{Siegel\,-Zwiebach equations}
\label{BeBo:SZeqs}

The Bargmann\,-Wigner programme for finite-component representations in Minkowski spacetime of any dimension 
$D>3$ was completed for massless helicity representations by Siegel and Zwiebach in \cite{Siegel:1986zi} and generalised to massive representations in Siegel's lecture notes \cite{Siegel:9912205}. Only the massless representations will be reviewed here since the case of massive representations follows by dimensional reduction, as mentioned in the subsection \ref{BeBo:massive}.

\newpage
\vspace{3mm}\textit{Siegel\,-Zwiebach equations}

\vspace{1mm}The main idea behind these equations is the covariantisation of the condition that the ``translation'' generators $\pi_n$ of the massless little group $IO(D-2)$ must act trivially on physical states of the helicity representations (cf. Subsections \ref{BeBo:orbit}-\ref{BeBo:classification}). Let us rewind the procedure initiated in Subsection \ref{BeBo:helicityreps}: 

\textbf{Steps 1 and 2.} These first steps are identical to the case of Bargmann\,-Wigner equations, \textit{i.e.} the wave function is a field strength ${\cal K}(p,d_{_{I}}x)$ 
taking values in an irrep of the Lorentz group $O(D-1,1)$ labeled by the Young diagram ${\overline{\lambda}}$.

\textbf{Step 3.} The generators of the Lorentz algebra $\mathfrak{so}(D-1,1)$ can be decomposed as the sum $M_{\mu\nu}=L_{\mu\nu}+S_{\mu\nu}$ of the ``orbital'' part (transforming the positions or momenta) and the ``spin'' part (transforming the irrep labeled by the Young diagram ${\overline{\lambda}}$),
\begin{align}
L_{\mu\nu}\,&=\,-i\,\Big(p_{\mu}\frac{\partial}{\partial p^{\nu}} - p_{\nu}\frac{\partial}{\partial p^{\mu}} \Big)\;,
\quad 
S_{\mu\nu}\,=\,-i\,\Big(d_{_I} x_{\mu}\frac{\partial}{\partial (dx_{_I}^{\nu})} - 
d_{_I} x_{\nu}\frac{\partial}{\partial (dx_{_I}^{\mu})}\Big)\,. 
\end{align}
The Siegel\,-Zwiebach equations for $s\neq 0$ take the simple form
\begin{equation}
(\,p^\mu S_{\mu\nu}\,-\,i\,s\,p_\nu\,)\,{\cal K}\,=\,0\,.
\label{BeBo:SZequations}
\end{equation}
They imply the massless Klein-Gordon equation $p^2{\cal K}=0$ (since $s\neq 0$). In fact, one can check that the quadratic and quartic Casimir operators both vanish as a consequence of \eqref{BeBo:SZequations}.\footnote{In order to check that the quartic Casmir operator acts trivially, it useful to notice that $M_{\mu\nu}$ can be replaced everywhere by $S_{\mu\nu}$ inside the definition \eqref{BeBo:QuartiC}. In $D=4$, this property is obvious in terms of the Pauli-Lubanski vector.} Notice that a similar ``spin-enslaving'' relation, leading to 
\eqref{BeBo:SZequations}, was recently given in \cite{Kuzenko:2020ayk}. 

\textbf{Step 4.} In the light-cone frame (see Section
\ref{BeBo:Lorentzgr}) where the components of the momentum are $p_{\mu}=(p_{-},0,0,\ldots, 0)$ with $p_-\neq 0\,$,
the system \eqref{BeBo:SZequations} of equations splits as
\begin{equation}
\pi_n\,{\cal K}\,=\,0\;,\quad(S_{+-}\,-\,i\,s){\cal K}\,=\,0\,,
\label{BeBo:spliSZequations}
\end{equation}
where $\pi_n:=p_- S_{+n}=p^\mu S_{\mu\, n}$ (with $n=1,2,\ldots,D-2$) are generators corresponding the ``translation'' subgroup $\mathbb{R}^{D-2}\subset IO(D-2)$ of the massless little group.\footnote{See Subsection
\ref{BeBo:orbit}. 
Note that $M_{+n}=S_{+n}$ and $M_{mn}=S_{mn}$ in this light-cone frame, since the corresponding orbital parts of the generators of the little group act trivially on the momentum.} 
On the one hand, the fact that these generators $\pi_n$ act trivially ensures that the massless 
representation is a helicity representation, \textit{i.e.} only the generators $S_{mn}$ of the 
rotations in the transverse plane act non-trivially. Moreover, the condition $\pi_n\,{\cal K}\,=\,0$ 
implies that the field strength ${\cal K}$ in the light-cone frame has a maximal (respectively, 
minimal) number of factors $d_{_I} x^-$ (respectively, $d_{_I}x^+$).\footnote{See \cite{Siegel:9912205} 
for an elegant derivation of these facts from \eqref{BeBo:spliSZequations}.} 
Therefore, the physical components of the field strength read ${\cal K}\,=\,(\prod_I
p_-d_{_I}x^-)\,\phi\,$, where $\phi$ depends neither on $d_{_I}x^-$ nor on $d_{_I}x^+$.
On the other hand, the eigenvalue $S_{+-}=is$ of the Lorentz generator 
\begin{align}
S_{+-}\,=\,-i\,\Big(d_{_I} x^+\frac{\partial}{\partial( d_{_I}x^+)}-d_{_I} x^-\frac{\partial}{\partial (d_{_I}x^-)}
\Big) 
\end{align}
implies that the Young diagram ${\overline{\lambda}}$ must have $s$ columns. 
This is because the operator $S_{+-}$ is a number operator (up to a coefficient $i$) 
for the total number of covariant indices $-$ minus the number of covariant 
indices $+\,$, and in every column of the field strength there is no index $+$ 
and one index $-\,$. 
The conclusion that is reached is the same as in Subsection \ref{BeBo:helicityreps}.

\vspace{3mm}\textit{Equivalence with Bargmann\,-Wigner equations}

\vspace{1mm}In fact, the Siegel\,-Zwiebach equations are equivalent to the Bargmann\,-Wigner equations reviewed in Subsection \ref{BeBo:helicityreps}. For instance, the closure and coclosure conditions \eqref{BeBo:closed} and \eqref{BeBo:coclosed} imply 
\eqref{BeBo:SZequations}. This follows from the identity
\begin{align}
p^\mu S_{\mu\nu}\,=&\,-i\,p^\mu\Big(d_{_I} x_{\mu}
\frac{\partial}{\partial (dx_{_I}^{\nu})}\,-\,d_{_I} x_{\nu}
\frac{\partial}{\partial (dx_{_I}^{\mu})}\Big)
\nonumber \\
\,=&
\,-i\,\Big(p_\mu d_{_I} x^\mu\Big)
\frac{\partial}{\partial (dx_{_I}^{\nu})}\,-\,d_{_I} x_{\nu}\Big(p^\mu\frac{\partial}{\partial (dx_{_I}^{\mu})}\Big)\,.
\end{align}
In the last term, one recognises between the parentheses the divergence operator 
acting on the column $I\,$, which gives zero because of the coclosure condition 
\eqref{BeBo:coclosed}. 
As for the first term on the right-hand side of the above equation, one can rewrite it as 
\begin{equation}
\,-i\,\Big(p_\mu d_{_I} x^\mu\Big)
\frac{\partial}{\partial (dx_{_I}^{\nu})} = \,-i\, 
\frac{\partial}{\partial (dx_{_I}^{\nu})} \Big(p_\mu \,d_{_I} x^\mu\Big)
\,-i\, p_{\mu}\Big[ dx_{_I}^{\mu} \,,\, \frac{\partial}{\partial (dx_{_I}^{\nu})} \Big] .
\end{equation}
The first term on the right-hand side gives zero on the field strength because of the 
closure relation \eqref{BeBo:closed}, while the last term gives $+i\,s\,p_{\nu}\,$ 
because of the commutation relations 
$\Big[ dx_{_I}^{\mu} , \frac{\partial}{\partial (dx_{_I}^{\nu})} \Big] 
= - s\,\delta^{\mu}_{\nu}\,$.

The covariant proof that the Siegel\,-Zwiebach equations imply Bargmann\,-Wigner equations is more cumbersome and will not be presented here. Anyway, this equivalence is guaranteed from the light-cone frame analysis.

\section{Bargmann\,-Wigner programme in three dimensions}
\label{App:A}

In this appendix we review results obtained in the literature concerning the Wigner and
Bargmann\,-Wigner programmes in Minkowski spacetime of dimension $D=2+1\,$. 
The former programme was achieved in \cite{BeBo:Binegar} along the lines of the seminal
paper \cite{BeBo:Wigner39} by Wigner.

There are four classes of UIRs of the Poincar\'e group $ISO(2,1)^{\uparrow}\,$:
\begin{itemize}
\item[1)]  
Zero-momentum representations, labeled by the eigenvalue $c\in\mathbb{R}$ of the quadratic Casimir operator ${\cal C}_{2}[\mathfrak{so}(2,1)]$ of the Lorentz algebra\,;\footnote{\label{FT}Strictly speaking, the 
principal and complementary series are labeled by two real parameters, not only by the value of the Casimir operator.}
\item[2)]  
Massive representations, labeled by mass $m>0$ and spin $s\in \mathbb{R}$\,;
\item[3)] 
Massless representations:
\begin{itemize}
	\item[1.] helicity representations, either single-valued (bosonic) or double-valued (fermionic);
	\item[2.] 
 infinite-spin representations, labeled by a dimensionful parameter $\mu>0\,$;
\end{itemize}
\item[4)] Tachyonic representations, labeled by a dimensionful parameter $m>0$ 
 and by a dimensionless parameter $s\in \mathbb{R}$ (the analogue of spin).
\end{itemize} 
In what follows, we briefly summarise exhaustive results on the completion of the Bargmann\,-Wigner programme in $D=2+1$ dimensions for the four classes of UIRs listed above.  

\subsection{Zero momentum representations}

Effectively, the zero momentum representation of the Poincar\'e group $ISO(2,1)^{\uparrow}$ are UIRs of the Lorentz subgroup $SO(2,1)^{\uparrow}\,$. The latter were classified in \cite{BeBo:Bargmann}. 
We also refer the reader to \cite{BaFr} for a physicist-friendly classification of the irreps of the Lorentz group $SO(2,1)^{\uparrow}\,$.

We will not repeat these well-known results here. For the purpose of the Bargmann\,-Wigner programme, it is enough to know that the UIRs of $SO(2,1)^{\uparrow}\,$ are labeled by the real eigenvalue of the quadratic Casimir operator ${\cal C}_{2}[\mathfrak{so}(2,1)]$ of the Lorentz algebra (and another real parameter for the 
principal and complementary series, cf. Footnote \ref{FT}). 
Since the momentum is vanishing, the states span a constant field 
$\psi$ on Minkowski spacetime taking values in these UIRs of the Lorentz group $SO(2,1)^{\uparrow}\,$.
A relativistic equation is then $\big(\,{\cal C}_{2}[\mathfrak{so}(2,1)]-c\,\big)\psi=0\,$, 
which asserts that the states $\psi$ are eigenvectors of the Casimir operator with eigenvalue $c\in\mathbb{R}\,$.  

\subsection{Massive representations}

Consider a massive representation labeled by mass $m>0$ and spin $s\in \mathbb{R}$\,.
   
\subsubsection{(Half-)integer spins}

For integer spin $s\in\mathbb{N}$, the
Klein-Gordon equation \eqref{BeBo:KGequ} together with the tracelessness condition \eqref{BeBo:tracelessym} 
and the transversality condition \eqref{BeBo:transvmassivec} 
for a totally symmetric tensor $\varphi_{\mu_{1}\ldots\mu_{s}}$ provide relativistic field equations whose positive-energy 
solutions represent the corresponding UIR. 
Equivalently, for non-vanishing integer spin $s\in\mathbb{N}_0\,$, 
they can be summarised by the following set of equations:
\begin{equation}
\eta^{\mu_{1}\mu_{2}}\,\varphi_{\mu_{1}\ldots \mu_{s}} = 0 
 \;,\quad
m\,\varphi_{\mu_{1}\ldots \mu_{s}} \pm i\epsilon_{\mu_{1}\nu\rho}\,
p^{\nu}\varphi^{\rho}{}_{\mu_{2}\ldots\mu_{s}} = 0\;.
\label{FP3}
\end{equation}
where we take $\epsilon_{012}=-1\,$. For early references, see 
\cite{Plyushchay:1990rt,BeBo:Jackiw:1990ka}. 
These equations are explicitly written in \cite{Tyutin:1997yn} and can 
be found in spinor notation in \cite{Gorbunov:1996ed}.
Notice that the transversality condition \eqref{BeBo:transvmassivec} directly follows from 
the second equation in \eqref{FP3}. 
Moreover, note that there is no need to explicitly symmetrize the last 
equation in its free indices when the tracelessness and transversality conditions hold true. 
In turn, the Klein-Gordon equation follows from repeated application of the second equation in \eqref{FP3}. 
The two possible signs in the last equation stand for the two possible values 
$\pm s$ of the ``helicity'' of the massive particle. 
This system of equations can be generalized to AdS$_{3}\,$;  
see \cite{Boulanger:2014vya} for a classification.

\subsubsection{Fractional spins}

In the case of the massive UIRs
 where the real number $s$ is neither integer nor half-integer (``fractional spin'', see e.g. 
 \cite{Forte:1990hd} for a review), 
 one should stress that although the number of physical components is one (the UIRs of the 
 massive little group $SO(2)$ are one-dimensional since this group is Abelian) nevertheless 
 their corresponding covariant description necessarily involve relativistic field equations with 
 an infinite number of components (since there are no finite-dimensional irreps of the Lorentz 
 group $SO(2,1)^\uparrow$ with such values of the spin).  

The positive-energy solutions to the system of the four equations
\eqref{BeBo:KGequ}, \eqref{BeBo:masscalar}, \eqref{BeBo:tracelessym}, \eqref{BeBo:transversalityequ} 
formally describe a massive UIR of mass $m$ and spin $s\in\mathbb{R}$ (as can be checked by computing 
the value of the quartic Casimir operator). Note that the field $\Phi(p,u)$ is not polynomial in the auxilliary vector $u^\mu$ when $s\notin\mathbb{N}$. Finding a suitable functional space is a subtle issue that we will not attempt to address. In fact, the construction of manifestly $IO(2,1)$-covariant field equations proved to be a rather difficult task.

Several approaches have been followed in the literature. 
We refer to reader to \cite{Forte:1990hd} and the introduction of the paper \cite{Horvathy:2010vm} 
for reviews; see also \cite{BeBo:Jackiw:1990ka,Plyushchay:1990rt,Gorbunov:1996ed}.
In the following, we will review the results obtained in \cite{Cortes:1992fa} for the linear 
relativistic equations whose positive-energy solutions span the massives UIRs where
the spin $s$ is neither integer nor half-integer. 

The Cortes-Plyushchay equations proposed in \cite{Cortes:1992fa} read\footnote{One 
can show that the operator $V_{\mu}$ can be obtained by the dimensional reduction 
of the Siegel\,-Zwiebach massless operator in \eqref{BeBo:SZequations}.} 
\begin{equation}
V_{\mu}\psi = 0\;,\qquad V_{\mu}:= s\,P_{\mu} 
- i\, \epsilon_{\mu\nu\lambda}P^{\nu}\widetilde{M}^{\lambda} 
+ m\, \widetilde{M}_{\mu}\;,
\label{CP}
\end{equation}
where the three operators $\widetilde{M}_{\mu}:=\frac{1}{2}\,\epsilon_{\mu\nu\rho}\,M^{\nu\rho}$ 
generate the 
$\mathfrak{so}(2,1)$ Lorentz algebra in $D=2+1$ dimensions (
$i\,[\widetilde{M}_{\mu},\widetilde{M}_{\nu}] = \epsilon_{\mu\nu\rho}\widetilde{M}^{\rho}\,$), 
so that the quadratic Casimir \eqref{BeBo:QuadratiC} is equal to
${\cal C}_{2}[\mathfrak{so}(2,1)]=-\widetilde{M}_{\mu}\widetilde{M}^{\mu}$. 
In the above equations \eqref{CP}, the real number $s$ is assumed to be nonzero. 
Contracting the above equations with $\widetilde{M}^{\mu}\,$, $P^{\mu}$  and 
$
\epsilon^{\mu\nu\lambda}P_{\mu}\widetilde{M}_{\lambda}\,$ produces the following three equations 
\begin{equation}
\big(\, (s-1)\, W  + m \,\widetilde{M}^2\,\big) \psi = 0\,,~ 
( s\, P^{2} + m \, W)\psi = 0\,,~
\big( P^{2}\widetilde{M}^2 + W (m-W)\,\big)\psi = 0\,, 
\end{equation}
where the scalar $W:=P^{\mu}\widetilde{M}_{\mu}$ is, in three spacetime dimension, the analogue of the Pauli-Lubanski vector.
Since by assumption both $s$ and $m$ are non-zero, these three equations are equivalent to 
\begin{equation}
\big(m^{2}\widetilde{M}^2-s(s-1)P^{2}\big)\psi = 0\;,~
( s\, P^{2} + m \, W )\psi = 0\;,~
P^{2}(P^{2}+m^{2})\psi = 0\;.
\end{equation}
If one discards the trivial representation of the Poincar\'e group where $P_{\mu}=0=\widetilde{M}_{\mu}\,$, 
one gets the following three equations:  
\begin{equation}
\big(\,\widetilde{M}^2+s(s-1)\,\big)\psi = 0\;,\quad
(W -s\,m)\psi = 0\;,\quad
(P^{2}+m^{2})\psi = 0\;,
\label{BeBo:Casimirs}
\end{equation}
the last two being the Pauli-Lubanski condition and the Klein-Gordon equation, whereas the first sets the quadratic Casimir of the Lorentz group to ${\cal C}_{2}[\mathfrak{so}(2,1)]=s(s-1)$, which indicates that the field $\psi$ takes value in an irrep of the Lorentz group labeled by $s\,$. The positive-energy 
solutions of the above field equations \eqref{BeBo:Casimirs} transform in the UIR of mass $m$ and spin $s$.
More directly, in the Lorentz frame where $p^{\mu}=(m,0,0)\,$, the
Cortes-Plyushchay equations \eqref{CP} yield
\begin{equation}
(\widetilde{M}_{0}-
s)\psi = 0\;,\qquad (\widetilde{M}_{1}-i 
 \widetilde{M}_{2})\psi = 0\;. 
\end{equation}
If one takes $L_\pm:=\widetilde{M}_{1}\pm i \widetilde{M}_{2}$ as raising/lowering operators of the Lorentz algebra $\mathfrak{so}(2,1)$, then
these equations assert that the state of momentum $p^{\mu}=(
 m,0,0)$ is a lowest-weight state of $\mathfrak{so}(2,1)$.
This implies that the positive-energy solutions are fields taking values in a representation of the Lorentz algebra bounded from below.
For $s\notin\frac12\mathbb{N}$\,, one concludes that the field $\psi$ 
takes values in an infinite-dimensional UIR of the Lorentz algebra $\mathfrak{so}(2,1)$ belonging 
to the discrete series.

The cases with $s=-j<0$, where $j\in\frac12\mathbb{N}$ is a non-vanishing (half)integer, correspond 
to the non-unitary spin-$j$ irreducible representations of the Lorentz algebra 
$\mathfrak{so}(2,1)\,$ with quadratic Casimir ${\cal C}_2[\mathfrak{so}(2,1)]=j(j+1)\,$, in which case the Cortes-Plyushchay equation propagates the massive fields with (half)integer spins discussed around \eqref{FP3}. 

Manifest covariance groups the three components of the equations as the components of a vector.
This being said, let us mention that only two of the three equations \eqref{CP} are enough to produce the third one.   
These equations are integrable in the sense that 
the commutator $[V_{\mu},V_{\nu}]\psi$ vanishes on a field $\psi$ solution of \eqref{CP}. 
We refer to \cite{Horvathy:2010vm} for an extended discussion of these equations.

\subsection{Massless representations}

The massless little group in $D=2+1$ spacetime dimensions is 
$ISO(1)\cong\mathbb R$ that is abelian, hence massless UIRs are 
one-dimensional and labeled by a single real parameter $\mu\in\mathbb R\,$. 
Therefore, all massless UIRs of the Poincar\'e group $ISO(2,1)^{\uparrow}$ have a single physical component. Nevertheless, we will stick to the distinction ``helicity'' vs ``infinite-spin'' representations.

\subsubsection{Helicity representations}

The helicity representations correspond to the particular case $\mu=0\,$. 
Two case arises whether the representation of the Lorentz group $SO(2,1)^{\uparrow}$ 
is either single or double valued: 
the ``helicity'' is effectively zero or one-half,
which corresponds to the fact that a massless field in three spacetime dimensions 
can always be dualized to 
a massless scalar or a Dirac spinor, as will be reviewed now.
The manifestly covariant field equations are similar to those for the massless helicity cases 
in $D>3$ studied above, except that only symmetric (spinor-)tensor gauge fields 
$\varphi_{\mu_{1}\ldots\mu_{s}}=\varphi_{(\mu_{1}\ldots\mu_{s})}$
are allowed (the spinor index is not written). Equivalently, only field strengths ${\cal K}_{\mu_1\nu_1\,\mid\ldots\,\mid\,\mu_s\nu_s}
$ labeled by rectangular two-row Young diagrams are allowed. Moreover,
higher (gamma-)traces of   
those field strengths must be set to zero. 
Indeed, if in three dimensions one were to set to zero the single (gamma-)trace of the 
field strength ${\cal K}
\,$, 
one would obtain that the field strength itself should vanish on-shell, 
resulting in the absence of propagating degrees of freedom.
More precisely, upon Hodge-dualizing the $s$ pairs of antisymmetric indices of the spin-$s$ field strength 
one obtains a totally symmetric (spinor-)tensor 
$$\widetilde{{\cal K}}_{\mu_{1}\ldots\mu_{s}}
\,:=\,\frac1{2^s}\,\epsilon_{\mu_1\nu_1\rho_1}\cdots\epsilon_{\mu_s\nu_s\rho_s}\,{\cal K}^{\nu_1\rho_1\,\mid\ldots\,\mid\,\nu_s\rho_s}
\,,$$ 
where the latter (spinor-)tensor is completely symmetric 
in its spacetime indices.  

The closure and coclosure conditions on the field strength $\cal K$ are
equivalent to coclosure and closure condition on its dual:
\begin{equation}
\partial^{\mu_{1}}\widetilde{{\cal K}}_{\mu_{1}\mu_2\ldots\mu_{s}}=0\;,\qquad 
\partial_{\mu}\widetilde{{\cal K}}_{\nu\rho_1\ldots\rho_{s-1}} - 
\partial_{\nu}\widetilde{{\cal K}}_{\mu\rho_1\ldots\rho_{s-1}} =0\;.
\end{equation}
The field strength $ \widetilde{K}$ begin closed, it is exact:
\begin{equation}
\widetilde{{\cal K}}_{\mu_{1}\ldots\mu_{s}} =p_{\mu_{1}}\ldots  p_{\mu_{s}}\phi\;,
\label{BeBo:phi}
\end{equation}
where $\phi$ is a (spinor) scalar. 

The higher-trace equations on the field strength $\cal K$ for a propagating, 
massless helicity representation in three dimensions, are then for bosons
\begin{equation}
\eta^{\mu_{1}\mu_{2}}\,\widetilde{{\cal K}}_{\mu_{1}\mu_{2}\mu_{3}\ldots\mu_{s}}=0\;,
\qquad s>1\;,
\label{BW3b}
\end{equation} 
with the usual massless Klein-Gordon and Maxwell equations for $s=0$ and $1$, 
respectively, and for fermions
\begin{equation}
\gamma^{\mu}\widetilde{\cal K}_{\mu\nu_{2}\ldots\nu_{s}}=0
\label{BW3f}
\end{equation}
for the spin $s+\tfrac{1}{2}>\frac{1}{2}\,$ cases; 
the spin-$\tfrac{1}{2}\,$ case being of course given by 
$\gamma^{\mu}\partial_{\mu}\phi = 0\,$, where again, 
the spinor indices are not written and the three $\gamma^{\mu}$ matrices 
are three Dirac (in fact Pauli) matrices in $D=2+1$ dimensions.

The conclusion is that all these descriptions of bosonic (respectively, fermionic) massless fields are dual to each others, 
for all (half-)integer values of the ``spin'' $s$, in accordance with fact that the positive-energy solutions of the above 
Bargmann\,-Wigner equations \eqref{BW3b} (respectively, \eqref{BW3f}, for fermions) carry a single (respectively, double) valued helicity representations of the Poincar\'e group $ISO(2,1)^{\uparrow}$. Concretely, these fields are dual a scalar (or spinor) field. More explicitly, the on-shell duality relation between the gauge fields 
$\varphi_{\mu_{1}\ldots\mu_{s}}$, the field strengths ${\cal K}_{\mu_1\nu_1\,\mid\ldots\,\mid\,\mu_s\nu_s}$ 
and the massless scalar (or spinor) field $\phi$ is \eqref{BeBo:phi}.

\subsubsection{Infinite spin representations}

The positive-energy solutions of the Wigner equations \eqref{BeBo:epsilon3}-\eqref{BeBo:epsilon2}, reviewed in Subsection \ref{BeBo:infspin},
 transform in the massless infinite-spin representation of the Poincar\'e group $ISO(2,1)^{\uparrow}\,$, labeled by $\mu>0$. The paper \cite{Schuster:2014xja} provided an extensive discussion of massless infinite-spin particles in $D=2+1$ dimensions.

\subsection{Tachyonic representations}

Finally, in order to be exhaustive, we end this section by mentioning that the relativistic equations \eqref{BeBo:tachyo1}-\eqref{homogen} 
provide an exhaustive solution of the Bargmann\,-Wigner programme in the tachyonic case. Indeed, the little group $SO(1,1)$ of a spacelike momenta in $D=2+1$ dimensions is Abelian, thus its UIRS of are labeled by a single parameter $s\in\mathbb{R}\,$. 

\end{appendix}





\end{document}